\documentclass[twocolumn,numbook,nospthms,natbib]{svjour3}          
\smartqed  
 \usepackage{graphicx}
\usepackage{amsmath,amssymb,amsthm} 
\usepackage{enumerate,epstopdf}
\usepackage{color}
\usepackage{pdfsync}
\usepackage{float}
\usepackage{graphicx}
\usepackage{xspace}
\usepackage{algpseudocode}
\usepackage{algorithmicx}
\usepackage{algorithm}

\def\cB{{\mathcal B}}

\def\cF{{\mathcal F}}
\def\sX{{\mathsf X}}
\def\cM{{\mathcal M}}
\def\cU{{\mathcal U}}

\def\cP{{\mathcal P}}
\def\bR{{\mathbb R}}

\def\bE{{\mathbb E}}

\def\bN{{\mathbb N}}

\def\NPDF{{\mathcal N}}

\def\f0{{\mathbf 0}}

\def\md{{\mbox d}}
\def\rw{\rightarrow}
\def\qed{$\Box$}
\def\sfp{{\sf p}}

\newtheorem{thm}{Theorem}
\newtheorem{defn}{Definition}
\newtheorem{assumption}{Assumption}
\newtheorem{lem}{Lemma}
\newtheorem{prop}{Proposition}

\begin{document}

\title{Nudging the particle filter\thanks{This work was partially supported by \textit{Ministerio de Econom\'ia y Competitividad} of Spain (TEC2015-69868-C2-1-R ADVENTURE), the Office of Naval Research Global (N62909-15-1-2011), and the regional government of Madrid (program CASICAM-CM S2013/ICE-2845).}
}


\author{\"Omer Deniz Akyildiz         \and
        Joaqu\'in M\'iguez 
}


\institute{\"O.~D.~Akyildiz \at
              {University of Warwick}\\
              The Alan Turing Institute\\
              \email{omer.akyildiz@warwick.ac.uk}           
           \and
           J.~M\'iguez \at
              Universidad Carlos III de Madrid \& Instituto de Investigaci\'on Sanitaria Gregorio Mara\~n\'on\\
}

\date{Received: date / Accepted: date}

\maketitle

\begin{abstract}
We investigate a new sampling scheme aimed at improving the performance of particle filters whenever (a) there is a significant mismatch between the assumed model dynamics and the actual system, or (b) the posterior probability tends to concentrate in relatively small regions of the state space. The proposed scheme pushes some particles towards specific regions where the likelihood is expected to be high, an operation known as \textit{nudging} in the geophysics literature. We re-interpret nudging in a form applicable to any particle filtering scheme, as it does not involve any changes in the rest of the algorithm. Since the particles are modified, but the importance weights do not account for this modification, the use of nudging leads to additional bias in the resulting estimators. However, we prove analytically that nudged particle filters can still attain asymptotic convergence with the same error rates as conventional particle methods. Simple analysis also yields an alternative interpretation of the nudging operation that explains its robustness to model errors. Finally, we show numerical results that illustrate the improvements that can be attained using the proposed scheme. In particular, we present nonlinear tracking examples with synthetic data and a model inference example using real-world financial data.  
\end{abstract}

\section{Introduction}\label{secIntro}

\subsection{Background}
State-space models (SSMs) are ubiquitous in many fields of science and engineering, including weather forecasting, mathematical finance, target tracking, machine learning, population dynamics, etc., where inferring the states of dynamical systems from data plays a key role.

A SSM comprises a pair of stochastic processes $(x_t)_{t\geq 0}$ and $(y_t)_{t\geq 1}$ called \textit{signal process} and \textit{observation process}, respectively. The conditional relations between these processes are defined with a transition and an observation model (also called {\em likelihood} model) where observations are conditionally independent given the signal process, and the latter is itself a Markov process. Given an observation sequence, $y_{1:t}$, the filtering problem in SSMs consists in the estimation of expectations with respect to the posterior probability distribution of the hidden states, conditional on $y_{1:t}$, which is also referred to as the filtering distribution.

Apart from a few special cases, neither the filtering distribution nor the integrals (or expectations) with respect to it can be computed exactly; hence, one needs to resort to numerical approximations of these quantities. Particle filters (PFs) have been a classical choice for this task since their introduction by \citet{gordon1993novel}; see also \citet{kitagawa1996monte, liu1998sequential, doucet2000sequential, de2001introduction}. The PF constructs an empirical approximation of the posterior probability distribution via a set of Monte Carlo samples (usually termed {\em particles}) which are modified or killed sequentially as more data are taken into account. These samples are then used to estimate the relevant expectations. The original form of the PF, often referred to as the bootstrap particle filter (BPF), has received significant attention due to its efficiency in a variety of problems, its intuitive appeal and its straightforward implementation. A large body of theoretical work concerning the BPF has also been compiled. For example, it has been proved that the expectations with respect to the empirical measures constructed by the BPF converge to the expectations with respect to the true posterior distributions when the number of particles is large enough \citep{del1999central, chopin2004central, kunsch2005recursive, douc2007limit} or that they converge uniformly over time under additional assumptions related to the stability of the true distributions \citep{del2001stability, del2004feynman}.

{
Despite the success of PFs in relatively low dimensional settings, their use has been regarded impractical in models where $(x_t)_{t\geq 0}$ and $(y_t)_{t\geq 1}$ are sequences of high-dimensional random variables. 
In such scenarios, standard PFs have been shown to \textit{collapse} \citep{bengtsson2008curse, snyder2008obstacles}. This problem has received significant attention from the data assimilation community. The high-dimensional models which are common in meteorology and other fields of Geophysics are often dealt with via an operation called \textit{nudging} \citep{hoke1976initialization,malanotte1986data,malanotte1988data,zou1992optimal}.} Within the particle filtering context, nudging can be defined as a transformation of the particles, which are pushed towards the observations using some observation-dependent map \citep{van2009particle,van2010nonlinear,ades2013exploration,ades2015equivalent}. If the dimensions of the observations and the hidden states are different, which is often the case, a gain matrix is computed in order to perform the nudging operation. In \citet{van2009particle,van2010nonlinear,ades2013exploration,ades2015equivalent}, nudging is performed after the sampling step of the particle filter. The importance weights are then computed accordingly, so that they remain proper. Hence, nudging in this version amounts to a sophisticated choice of the importance function that generates the particles. It has been shown (numerically) that the schemes proposed by \citet{van2009particle,van2010nonlinear,ades2013exploration,ades2015equivalent} can track high-dimensional systems with a low number of particles. However, generating samples from the nudged proposal requires costly computations for each particle and the evaluation of weights becomes heavier as well. It is also unclear how to apply existing nudging schemes when non-Gaussianity and nontrivial nonlinearities are present in the observation model.

A related class of algorithms includes the so-called implicit particle filters (IPFs) \citep{chorin2009implicit,chorin2010implicit,atkins2013implicit}. Similar to nudging schemes, IPFs rely on the principle of pushing particles to high-probability regions in order to prevent the collapse of the filter in high-dimensional state spaces. In a typical IPF, the region where particles should be generated is determined by solving an algebraic equation. This equation is model dependent, yet it can be solved for a variety of different cases (general procedures for finding solutions are given by \citet{chorin2009implicit} and \citet{chorin2010implicit}). The fundamental principle underlying IPFs, moving the particles towards high-probability regions, is similar to nudging. Note, however, that unlike IPFs, nudging-based methods are not designed to {\em guarantee} that the resulting particles land on high-probability regions; it can be the case that nudged particles are moved to relatively low probability regions (at least occasionally). Since an IPF requires the solution of a model-dependent algebraic equation for every particle, it can be computationally costly, similar to the nudging methods by \citet{van2009particle,van2010nonlinear,ades2013exploration,ades2015equivalent}.  Moreover, it is not straightforward to derive the map for the translation of particles in general models, hence the applicability of IPFs depends heavily on the specific model at hand.
%
%

\subsection{Contribution}

In this work, we propose a modification of the PF, termed \textit{the nudged particle filter} (NuPF) and assess its performance in high dimensional settings and with misspecified models. Although we use the same idea for nudging that is presented in the literature, our algorithm has subtle but crucial differences, as summarized below.

\begin{itemize}
\item First, we define the nudging step not just as a relaxation step towards observations but as a step that strictly increases the likelihood of a subset of particles. This definition paves the way for different nudging schemes, such as using the gradients of likelihoods or employing random search schemes to move around the state-space. In particular, classical nudging (relaxation) operations arise as a special case of nudging using gradients when the likelihood is assumed to be Gaussian. Compared to IPFs, the nudging operation we propose is easier to implement as we only demand the likelihood to increase (rather than the posterior density). Indeed, nudging operators can be implemented in relatively straightforward forms, without the need to solve model-dependent equations.
\item Second, unlike the other nudging based PFs, we do not correct the bias induced by the nudging operation during the weighting step. {Instead, we compute the weights in the same way they would be computed in a conventional (non-nudged) PF and the nudging step is devised to preserve the convergence rate of the PF, under mild standard assumptions, despite the bias. 
Moreover, computing biased weights  is usually faster than computing proper (unbiased) weights. Depending on the choice of nudging scheme, the proposed algorithm can have an almost negligible computational overhead compared to the conventional PF from which it is derived.}

\item {Finally, we show that a nudged PF for a given SSM (say $\cM_0$) is equivalent to a standard BPF running on a modified dynamical model (denoted $\cM_1$). In particular, model $\cM_1$ is endowed with the same likelihood function as $\cM_0$ but the transition kernel is observation-driven in order to match the nudging operation. As a consequence, the implicit model $\cM_1$ is ``adapted to the data'' and we have empirically found that, for any sufficiently long sequence $y_1, \ldots, y_t$, the evidence\footnote{{Given a data set $\{ y_1, \ldots, y_t \}$, the evidence in favour of a model $\cM$ is the joint probability density of $y_1, \cdots, y_t$ conditional on $\cM$, denoted $\sfp(y_{1:t}|\cM)$.}} \citep{Robert07} in favour of $\cM_1$ is greater than the evidence in favour of $\cM_0$.} We can show, for several examples, that this implicit adaptation to the data makes the NuPF robust to mismatches in the state equation of the SSM compared to conventional PFs. In particular, provided that the likelihoods are specified or calibrated reliably, we have found that NuPFs perform reliably under a certain amount of mismatch in the transition kernel of the SSM, while standard PFs degrade clearly in the same scenario.
\end{itemize}

In order to illustrate the contributions outlined above, we present the results of several computer experiments with both synthetic and real data. In the first example, we assess the performance of the NuPF when applied to a linear-Gaussian SSM. The aim of these computer simulations is to compare the estimation accuracy and the computational cost of the proposed scheme with several other competing algorithms, namely a standard BPF, a PF with optimal proposal function and a NuPF with proper weights. The fact that the underlying SSM is linear-Gaussian enables the computation of the optimal importance function (intractable in a general setting) and proper weights for the NuPF. We implement the latter scheme because of its similarity to standard nudging filters in the literature. This example shows that the NuPF suffers just from a slight performance degradation compared to the PF with optimal importance function or the NuPF with proper weights, while the latter two algorithms are computationally more demanding.  

The second and third examples are aimed at testing the robustness of the NuPF when there is a significant misspecification in the state equation of the SSM. This is helpful in real-world applications because practitioners often have more control over measurement systems, which determine the likelihood, than they have over the state dynamics. We present computer simulation results for a stochastic Lorenz 63 model and a maneuvering target tracking problem. 

In the fourth example, we present numerical results for a  stochastic Lorenz 96 model, in order to show how a relatively high-dimensional system can be tracked without a major increase of the computational effort compared to the standard BPF. For this set of computer simulations we have also compared the NuPF with the Ensemble Kalman filter (EnKF), which is the de facto choice for tackling this type of systems.

Let us remark that, for the two stochastic Lorenz systems, the Markov kernel in the SSM can be sampled in a relatively straightforward way, yet transition probability densities cannot be computed (as they involve a sequence of noise variables mapped by a composition of nonlinear functions). Therefore, computing proper weights for proposal functions other than the Markov kernel itself is, in general, not possible for these examples.

Finally, we demonstrate the practical use of the NuPF on a problem where a real dataset is used to fit a stochastic volatility model using either particle Markov chain Monte Carlo (pMCMC) \citep{andrieu2010particle} or nested particle filters \citep{Crisan18bernoulli}.

\subsection{Organisation}

The paper is structured as follows. After a brief note about notation, we describe the SSMs of interest and the BPF in Section~\ref{secSSMBPF}. Then in Section~\ref{secNuPF}, we outline the general algorithm and the specific nudging schemes we propose to use within the PF. We prove a convergence result in Section~\ref{secAnalysis} which shows that the new algorithm has the same asymptotic convergence rate as the BPF. We also provide an alternative interpretation of the nudging operation that explains its robustness in scenarios where there is a mismatch between the observed data and the assumed SSM. We discuss the computer simulation experiments in Section~\ref{secNumericalSims} and present results for real data in Section~\ref{secExperiments}. Finally, we make some concluding remarks in Section~\ref{secConclusions}.

%
\subsection{Notation}
We denote the set of real numbers as $\bR$, while $\bR^d = \bR \times \stackrel{d}{\cdots} \times \bR$ is the space of $d$-dimensional real vectors. We denote the set of positive integers with $\bN$ and the set of positive reals with $\bR_+$. We represent the state space with $\mathsf{X} \subset \bR^{d_x}$ and the observation space with $\mathsf{Y} \subset \bR^{d_y}$.

In order to denote sequences, we use the shorthand notation $x_{i_1:i_2} = \{x_{i_1},\ldots,x_{i_2}\}$. For sets of integers, we use $[n] = \{1,\ldots,n\}$. The  $p$-norm of a vector $x\in\bR^d$ is defined by $\|x\|_p = (x_1^p + \cdots + x_d^p)^{{1}/{p}}$. The $L_p$ norm of a random variable $z$ with probability density function (pdf) $p(z)$ is denoted $\|z\|_p = \left(\int |z|^p p(z) \md z\right)^{1/p}$, for $p\ge 1$. The Gaussian (normal) probability distribution with mean $m$ and covariance matrix $C$ is denoted $\NPDF(m,C)$. We denote the identity matrix of dimension $d$ with $I_d$.

The supremum norm of a real function $\varphi:\mathsf{X} \to \bR$ is denoted $\|\varphi\|_\infty = \sup_{x \in \mathsf{X}} |\varphi(x)|$. A function is bounded if $\| \varphi \|_\infty < \infty$ and we indicate the space of real bounded functions $\mathsf{X} \rw \bR$ as $B(\mathsf{X})$. The set of probability measures on $\sX$ is denoted $\cP(\sX)$, the Borel $\sigma$-algebra of subsets of $\mathsf{X}$ is denoted $\cB(\mathsf{X})$ and the integral of a function $\varphi:\mathsf{X}\rw\bR$ with respect to a measure $\mu$ on the measurable space $\left(\mathsf{X},\cB(\mathsf{X})\right)$ is denoted $(\varphi,\mu):=\int \varphi\md\mu$. The unit Dirac delta measure located at $x \in \bR^d$ is denoted $\delta_x(\md x)$. The Monte Carlo approximation of a measure $\mu$ constructed using $N$ samples is denoted as $\mu^N$. Given a Markov kernel $\tau(\mbox{d}x'|x)$ and a measure $\pi(\mbox{d}x)$, we define the notation $\xi(\mbox{d}x') = \tau \pi \triangleq \int \tau(\mbox{d}x'|x) \pi(\mbox{d}x)$.


\section{Background}\label{secSSMBPF}

%
\subsection{State space models}

We consider SSMs of the form
\begin{align}
x_0 &\sim \pi_0(\mbox{d} x_0), \label{prioreq} \\
x_t | x_{t-1} &\sim \tau_t(\mbox{d} x_t|x_{t-1}), \label{transitionEq} \\
y_t | x_t &\sim g_t(y_t|x_t), \quad t \in \bN, \label{observationEq}
\end{align}
where $x_t \in \mathsf{X}$ is the system state at time $t$, $y_t \in \mathsf{Y}$ is the $t$-th observation, the measure $\pi_0$ describes the prior probability distribution of the initial state, $\tau_t$ is a Markov transition kernel on $\mathsf{X}$, and $g_t(y_t|x_t)$ is the (possibly non-normalised) pdf of the observation $y_t$ conditional on the state $x_t$. We assume the observation sequence $\{ y_t \}_{t \in \bN_+}$ is arbitrary but fixed. Hence, it is convenient to think of the conditional pdf $g_t$ as a likelihood function and we write $g_t(x_t) := g_t(y_t|x_t)$ for conciseness.

We are interested in the sequence of posterior probability distributions of the states generated by the SSM. To be specific, at each time $t=1, 2, ...$ we aim at computing (or, at least, approximating) the probability measure $\pi_t$ which describes the probability distribution of the state $x_t$ conditional on the observation of the sequence $y_{1:t}$. When it exists, we use $\pi(x_t|y_{1:t})$ to denote the pdf of $x_t$ given $y_{1:t}$ with respect to the Lebesgue measure, i.e., $\pi_t(\md x_t) = \pi(x_t|y_{1:t})\md x_t$.

The measure $\pi_t$ is often termed the {\em optimal filter} at time $t$. It is closely related to the probability measure $\xi_t$, which describes the probability distribution of the state $x_t$ conditional on $y_{1:t-1}$ and it is, therefore, termed the {\em predictive} measure at time $t$. As for the case of the optimal filter, we use $\xi(x_t|y_{1:t-1})$ to denote the pdf, with respect to the Lebesgue measure, of $x_t$ given $y_{1:t-1}$.

%
\subsection{Bootstrap particle filter}

The BPF \citep{gordon1993novel} is a recursive algorithm that produces successive Monte Carlo approximations of $\xi_t$ and $\pi_t$ for $t=1, 2, ...$. The method can be outlined as shown in Algorithm \ref{AlgBootstrap}. 

\begin{algorithm}[hbt]
\begin{algorithmic}[1]
\caption{Bootstrap Particle Filter}\label{AlgBootstrap}
\State Generate the initial particle system $\{x_0^{(i)}\}_{i=1}^N$ by drawing $N$ times independently from the prior $\pi_0$.
\For{$t\geq 1$}
\State Sampling: draw $\bar{x}_t^{(i)} \sim \tau_t(\md x_t | x_{t-1}^{(i)})$ independently for every $i = 1, \ldots, N$.
\State Weighting: compute $w_t^{(i)} = {g_t(\bar{x}_t^{(i)})}/{\bar{Z}_t^N}$ for every $i = 1, \ldots, N$, where $\bar{Z}_t^N = \sum_{i=1}^N g_t(\bar{x}_t^{(i)})$.
\State Resampling: draw $x_t^{(i)}$, $i=1, ..., N$ from the discrete distribution $\sum_i w_t^{(i)} \delta_{\bar{x}_t^{(i)}}(\md x)$, independently for $i=1, ..., N$.
\EndFor
\end{algorithmic}
\end{algorithm}

After an initialization stage, where a set of independent and identically distributed (i.i.d.) samples from the prior are drawn, it consists of three recursive steps which can be depicted as,
\begin{equation}
\pi_{t-1}^N \underbrace{\to}_{\textnormal{sampling}} \xi_t^N \underbrace{\to}_{\textnormal{weighting}} \tilde{\pi}_t^N \underbrace{\to}_{\textnormal{resampling}} \pi_t^N.
\label{eqSchematicBPF}
\end{equation}
Given a Monte Carlo approximation $\pi_{t-1}^N = \frac{1}{N} \sum_{i=1}^N \delta_{x_{t-1}^{(i)}}$ computed at time $t-1$, the sampling step yields an approximation of the predictive measure $\xi_t$ of the form 
\begin{align*}
\xi_t^N = \frac{1}{N} \sum_{i=1}^N \delta_{\bar{x}_t^{(i)}}
\end{align*}
by propagating the {\em particles} $\{ x_{t-1}^{(i)} \}_{i=1}^N$ via the Markov kernel $\tau_t(\cdot|x_{t-1}^{(i)})$. The observation $y_t$ is assimilated via the importance weights $w_t^{(i)} \propto g_t(x_t^{(i)})$, to obtain the approximate filter
\begin{align*}
\tilde{\pi}_t^N = \sum_{i=1}^N w_t^{(i)} \delta_{\bar{x}_t^{(i)}},
\end{align*}
and the resampling step produces a set of un-weighted particles that completes the recursive loop and yields the approximation
\begin{align*}
\pi_t^N = \frac{1}{N} \sum_{i=1}^N \delta_{x_t^{(i)}}.
\end{align*}
The random measures $\xi_t^N$, $\tilde \pi_t^N$ and $\pi_t^N$ are commonly used to estimate {\em a posteriori} expectations conditional on the available observations. For example, if $\varphi$ is a function $\mathsf{X}\rw\bR$, then the expectation of the random variable $\varphi(x_t)$ conditional on $y_{1:t-1}$ is $\bE\left[ \varphi(x_t) | y_{1:t-1} \right] = (\varphi,\xi_t)$. The latter integral can be approximated using $\xi_t^N$, namely,
\begin{align*}
(\varphi,\xi_t) &= \int \varphi(x_t) \xi_t(\md x_t) \approx (\varphi,\xi_t^N) \\
&= \int \varphi(x_t)\xi_t^N(\md x_t) = \frac{1}{N} \sum_{i=1}^N \varphi(\bar x_t^{(i)}).
\end{align*}
Similarly, we can have estimators $(\varphi,\tilde \pi_t^N) \approx (\varphi,\pi_t)$ and $(\varphi,\pi_t^N) \approx (\varphi,\pi_t)$. Classical convergence results are usually proved for real bounded functions, e.g., if $\varphi \in B(\mathsf{X})$ then 
\begin{align*}
\lim_{N\rw\infty} | (\varphi,\pi_t) - (\varphi,\pi_t^N) | = 0 \quad \mbox{almost surely (a.s.)}
\end{align*}
under mild assumptions; see \citet{del2004feynman,crisan2009fundamentals} and references therein.

The BPF can be generalized by using arbitrary proposal pdf's $q_t(x_t|x_{t-1}^{(i)},y_t)$, possibly observation-dependent, instead of the Markov kernel $\tau_t(\cdot|x_{t-1}^{(i)})$ in order to generate the particles $\{ \bar x_t^{(i)} \}_{i=1}^N$ in the sampling step. This can lead to more efficient algorithms, but the weight computation has to account for the new proposal and we obtain \citep{doucet2000sequential}
\begin{align}
w_t^{(i)} \propto \frac{g_t(\bar{x}_t^{(i)}) \tau_t(\bar{x}_t^{(i)} | x_t^{(i)})}{q_t(\bar{x}_t^{(i)} | x_{t-1}^{(i)}, y_t)},
\label{eqComplicatedW}
\end{align}
which can be more costly to evaluate. This issue is related to the nudged PF to be introduced in Section \ref{secNuPF} below, which can be interpreted as a scheme to choose a certain observation-dependent proposal $q_t(x_t|x_{t-1}^{(i)},y_t)$. However, the new method does not require that the weights be computed as in \eqref{eqComplicatedW} in order to ensure convergence of the estimators.

\section{Nudged Particle Filter}\label{secNuPF}

%
\subsection{General algorithm}

Compared to the standard BPF, the nudged particle filter (NuPF) incorporates one additional step right after the sampling of the particles $\{ \bar x_t^{(i)} \}_{i=1}^N$ at time $t$. The schematic depiction of the BPF in \eqref{eqSchematicBPF} now becomes
\begin{align}
\pi_{t-1}^N \underbrace{\to}_{\textnormal{sampling}} \xi_t^N \underbrace{\to}_{\textnormal{nudging}} \tilde{\xi}_t^N \underbrace{\to}_{\textnormal{weighting}} \tilde{\pi}_t^N \underbrace{\to}_{\textnormal{resampling}} \pi_t^N,
\label{eqSchematicNuPF}
\end{align}
where the new {\em nudging step} intuitively consists in pushing a subset of the generated particles $\{ \bar x_t^{(i)} \}_{i=1}^N$ towards regions of the state space $\mathsf{X}$ where the likelihood function $g_t(x)$ takes higher values.

When considered jointly, the sampling and nudging steps in \eqref{eqSchematicNuPF} can be seen as sampling from a proposal distribution which is obtained by modifying the kernel $\tau_t(\cdot|x_{t-1})$ in a way that depends on the observation $y_t$. Indeed, this is the classical view of nudging in the literature \citep{van2009particle,van2010nonlinear,ades2013exploration,ades2015equivalent}. However, unlike in this classical approach, here the weighting step does not account for the effect of nudging. In the proposed NuPF, the weights are kept the same as in the original filter, $w_t^{(i)} \propto g_t(x_t^{(i)})$. In doing so, we save computations but, at the same time, introduce bias in the Monte Carlo estimators. One of the contributions of this paper is to show that this bias can be controlled using simple design rules for the nudging step, while practical performance can be improved at the same time.

In order to provide an explicit description of the NuPF, let us first state a definition for the nudging step.
\begin{defn} \label{defNudging} A nudging operator $\alpha_t^{y_t}: \mathsf{X} \to \mathsf{X}$ associated with the likelihood function $g_t(x)$ is a map such that
\begin{align}
\textnormal{if} \,\,\,\,\, x' = \alpha_t^{y_t}(x) \,\,\,\,\,\textnormal{then}\,\,\,\, g_t(x') \geq g_t(x)
\end{align}
for every $x,x'\in\mathsf{X}$.
\end{defn}
Intuitively, we define nudging herein as an operation that increases the likelihood. There are several ways in which this can be achieved and we discuss some examples in Sections \ref{secHowToChoose} and \ref{secHowToNudge}. The NuPF with nudging operator $\alpha_t^{y_t}:\mathsf{X}\rw\mathsf{X}$ is outlined in Algorithm \ref{Algnudged}. 

\begin{algorithm}[htb]
\begin{algorithmic}[1]
\caption{Nudged Particle Filter (NuPF)}\label{Algnudged}
\State Generate the initial particle system $\{x_0^{(i)}\}_{i=1}^N$ by drawing $N$ times independently from the prior $\pi_0$.
\For{$t\geq 1$}
\State Sampling: draw $\bar{x}_t^{(i)} \sim \tau_t(\md x_t | x_{t-1}^{(i)})$ independently for every $i = 1,\ldots, N$.
\State \textbf{Nudging}: choose a set of indices $\mathcal{I}_t \subset [N]$, then compute $\tilde{x}_t^{(i)} = \alpha_t^{y_t}(\bar{x}_{t}^{(i)})$ for every $i \in \mathcal{I}_t$. Keep $\tilde{x}_t^{(i)} = \bar x_t^{(i)}$ for every $i \in [N] \backslash \mathcal{I}_t$.
\State Weighting: compute $w_t^{(i)} = {g_t(\tilde{x}_t^{(i)})}/{\tilde{Z}_t^{N}}$ for every $i = 1,\ldots,N$, where $\tilde{Z}_t^N = \sum_{i=1}^N g(\tilde{x}_t^{(i)})$.
\State Resample: draw $x_t^{(i)}$ from $\sum_i w_t^{(i)} \delta_{\tilde{x}_t^{(i)}}(\md x)$ independently for $i=1, ..., N$.
\EndFor
\end{algorithmic}
\end{algorithm}

It can be seen that the nudging operation is implemented in two stages.
\begin{itemize}
\item First, we choose a set of indices $\mathcal{I}_t \subset [N]$ that identifies the particles to be nudged. Let $M=|\mathcal{I}_t|$ denote the number of elements in $\mathcal{I}_t$. We prove in Section \ref{secAnalysis} that keeping $M \le \mathcal{O}({\sqrt{N}})$ allows the NuPF to converge with the same error rates $\mathcal{O}({1}/{\sqrt{N}})$ as the BPF. In Section \ref{secHowToChoose} we discuss two simple methods to build $\mathcal{I}_t$ in practice.
\item Second, we choose an operator $\alpha_t^{y_t}$ that guarantees an increase of the likelihood of any particle. We discuss different implementations of $\alpha_t^{y_t}$ in Section \ref{secHowToNudge}.
\end{itemize}
We devote the rest of this section to a discussion of how these two steps can be implemented (in several ways).

%
\subsection{Selection of particles to be nudged} \label{secHowToChoose}

The set of indices $\mathcal{I}_t$, that identifies the particles to be nudged in Algorithm \ref{Algnudged}, can be constructed in several different ways, either random or deterministic. In this paper, we describe two simple random procedures with little computational overhead.  



\begin{itemize}
\item {\em Batch nudging:} Let the number of nudged particles $M$ be fixed. A simple way to construct $\mathcal{I}_t$ is to draw indices $i_1, i_2, \ldots, i_M$ uniformly from $[N]$ without replacement, and then let $\mathcal{I}_t = i_{1:M}$. We refer to this scheme as {\em batch nudging}, referring to selection of the indices at once. One advantage of this scheme is that the number of particles to be nudged, $M$, is deterministic and can be set a priori.  


\item {\em Independent nudging:} The size and the elements of $\mathcal{I}_t$ can also be selected randomly in a number of ways. Here, we have studied a procedure in which, for each index $i = 1, ..., N$, we assign $i \in \mathcal{I}_t$ with probability $\frac{M}{N}$. In this way, the actual cardinality $|\mathcal{I}_t|$ is random, but its expected value is exactly $M$. This procedure is particularly suitable for parallel implementations, since each index can be assigned to $\mathcal{I}_t$ (or not) at the same time as all others.

\end{itemize}

%

%
\subsection{How to nudge} \label{secHowToNudge}

The nudging step is aimed at increasing the likelihood of a subset of individual particles, namely those with indices contained in $\mathcal{I}_t$. Therefore, any map $\alpha_t^{y_t}:\mathsf{X}\rw\mathsf{X}$ such that $(g_t \circ \alpha_t^{y_t})(x) \ge g_t(x)$ when $x\in\mathsf{X}$ is a valid nudging operator. Typical procedures used for optimisation, such as gradient moves or random search schemes, can be easily adapted to implement (relatively) inexpensive nudging steps. Here we briefly describe a few of such techniques.


\begin{itemize}
\item {\em Gradient nudging:} If $g_t(x_t)$ is a differentiable function of $x_t$, one straightforward way to nudge particles is to take gradient steps. In Algorithm~\ref{gradientNudging} we show a simple procedure with one gradient step alone, and where {$\gamma_t>0$} is a step-size parameter and $\nabla_x g_t(x)$ denotes the vector of partial derivatives of $g_t$ with respect to the state variables, i.e., 
\begin{align*}
\nabla_{x_t} g_t = \left[
	\begin{array}{c}
	\frac{\partial g_t}{\partial x_{1,t}}\\
	\frac{\partial g_t}{\partial x_{2,t}}\\
	\vdots\\
	\frac{\partial g_t}{\partial x_{d_x,t}}\\
	\end{array}
\right] \quad \mbox{for} \quad x_t = \left[
	\begin{array}{c}
	x_{1,t}\\
	x_{2,t}\\
	\vdots\\
	x_{d_x,t}\\
	\end{array}
\right] \in \mathsf{X}.
\end{align*}
Algorithms can obviously be designed where nudging involves several gradient steps. In this work we limit our study to the single-step case, which is shown to be effective and keeps the computational overhead to a minimum. We also note that the performance of gradient nudging can be sensitive to the choice of the step-size parameters {$\gamma_t>0$}, which are, in turn, model dependent\footnote{{We have found, nevertheless, that fixed step-sizes (i.e., $\gamma_t=\gamma$ for all $t$) work well in practice for the examples of Sections \ref{secNumericalSims} and \ref{secExperiments}.}}. 

\item {\em Random nudging:} Gradient-free techniques inherited from the field of global optimisation can also be employed in order to {push} particles towards regions where they have higher likelihoods. A simple stochastic-search technique adapted to the nudging framework is shown in Algorithm~\ref{randomSearchNudging}. We hereafter refer to the latter scheme as random-search nudging.


\item {\em Model specific nudging:} Particles can also be nudged using the specific model information. For instance, in some applications the state vector $x_t$ can be split into two subvectors, $x_t^\textnormal{obs}$ and $x_t^\textnormal{unobs}$ (observed and unobserved, respectively), such that $g_t(x_t) = g_t(x_t^\textnormal{obs})$, i.e., the likelihood depends only on $x_t^\textnormal{obs}$ and not on $x_t^\textnormal{unobs}$. If the relationship between $x_t^\textnormal{obs}$ and $x_t^\textnormal{unobs}$ is tractable, one can first nudge $x_t^\textnormal{obs}$ in order to increase the likelihood and then modify $x_t^\textnormal{unobs}$ in order to keep it coherent with $x_t^\textnormal{obs}$. A typical example of this kind arises in object tracking problems, where positions and velocities have a special and simple physical relationship but usually only position variables are observed through a linear or nonlinear transformation. In this case, nudging would only affect the position variables. However, using these position variables, one can also nudge velocity variables with simple rules. We discuss this idea and show numerical results in Section~\ref{secNumericalSims}.

\end{itemize}

\begin{algorithm}[htb]
\begin{algorithmic}[1]
\caption{Gradient nudging}\label{gradientNudging}
\For{every $i \in \mathcal{I}_t$}
\begin{align*}
\tilde{x}_t^{(i)} = \bar{x}_t^{(i)} + \gamma_t \nabla_{x_t} g_t(\bar{x}_t^{(i)})
\end{align*}
\EndFor
\end{algorithmic}
\end{algorithm}

\begin{algorithm}[htb]
\begin{algorithmic}[1]
\caption{Random search nudging}\label{randomSearchNudging}
\Repeat
\State Generate $\tilde{x}_t^{(i)} = \bar{x}_t^{(i)} + \eta_t$ where $\eta_t \sim \NPDF(0,C)$ for some covariance matrix $C$.
\State If $g_t(\tilde{x}_t^{(i)}) > g_t(\bar{x}_t^{(i)})$ then keep $\tilde{x}_t^{(i)}$ , otherwise set $\tilde{x}_t^{(i)} = \bar{x}_t^{(i)}$.
\Until{the particle is nudged.}
\end{algorithmic}
\end{algorithm}

%
\subsection{Nudging general particle filters}

In this paper we limit our presentation to BPFs in order to focus on the key concepts of nudging and to ease presentation. It should be apparent, however, that nudging steps can be plugged into general PFs. More specifically, since the nudging step is algorithmically detached from the sampling and weighting steps, it can be easily used within any PF, even if it relies on different proposals and different weighting schemes. We leave for future work the investigation of the performance of nudging within widely used PFs, such as auxiliary particle filters (APFs) \citep{pitt1999filtering}. 


\section{Analysis}\label{secAnalysis}

The nudging step modifies the random generation of particles in a way that is not compensated by the importance weights. Therefore, we can expect nudging to introduce bias in the resulting estimators in general. However, in Section \ref{ssConvergenceLp} we prove that, as long as some basic guidelines are followed, the estimators of integrals with respect to the filtering measure $\pi_t$ and the predictive measure $\xi_t$ converge in $L_p$ as $N\rw\infty$ with the usual Monte Carlo rate $\mathcal{O}(1/\sqrt{N})$. The analysis is based on a simple induction argument and ensures the consistency of a broad class of estimators. In Section \ref{ssUniform} we briefly comment on the conditions needed to guarantee that convergence is attained uniformly over time. We do not provide a full proof, but this can be done by extending the classical arguments in \citet{del2001stability} or \citet{del2004feynman} and using the same treatment of the nudging step as in the induction proof of Section \ref{ssConvergenceLp}. Finally, in Section \ref{ssModelling}, we provide an interpretation of nudging in a scenario with modelling errors. In particular, we show that the NuPF can be seen as a {\em standard} BPF for a modified dynamical model which is ``a better fit'' for the available data than the original SSM.



%
\subsection{Convergence in $L_p$} \label{ssConvergenceLp}

The goal in this section is to provide theoretical guarantees of convergence for the NuPF under mild assumptions. First, we analyze a general NuPF (with arbitrary nudging operator $\alpha_t^{y_t}$ and an upper bound on the size $M$ of the index set $\mathcal{I}_t$) and then we provide a result for a NuPF with gradient nudging.

Before proceeding with the analysis, let us note that the NuPF produces several approximate measures, depending on the set of particles (and weights) used to construct them. After the sampling step, we have the random probability measure 
\begin{equation}
\xi_t^N = \frac{1}{N} \sum_{i=1}^N \delta_{\bar x_t^{(i)}},
\end{equation}
which converts into
\begin{equation}
\tilde \xi_t^N = \frac{1}{N} \sum_{i=1}^N \delta_{\tilde x_t^{(i)}}
\label{eqDef00}
\end{equation}
after nudging. Once the weights $w_t^{(i)}$ are computed, we obtain the approximate filter 
\begin{equation}
\tilde \pi_t^N = \sum_{i=1}^N w_t^{(i)} \delta_{\tilde x_t^{(i)}},
\end{equation}
which finally yields
\begin{equation}
\pi_t^N =  \frac{1}{N} \sum_{i=1}^N \delta_{x_t^{(i)}}
\label{eqDef01}
\end{equation} 
after the resampling step.

Similar to the BPF, the simple Assumption \ref{BoundedAssumption} stated next is sufficient for consistency and to obtain explicit error rates \citep{DelMoral00,Crisan02,miguez2013convergence} for the NuPF, as stated in Theorem \ref{ThmNudgingBound} below.

\begin{assumption}\label{BoundedAssumption} The likelihood function is positive and bounded, i.e.,
\begin{align*}
g_t(x_t) > 0 \,\,\,\,\textnormal{and}\,\,\,\,\|g_t\|_\infty = \sup_{x_t\in\mathsf{X}} |g_t(x_t)| < \infty
\end{align*}
for $t = 1,\ldots,T$.
\end{assumption}

\begin{thm}\label{ThmNudgingBound} 
Let $y_{1:T}$ be an arbitrary but fixed sequence of observations, with $T<\infty$, and choose any $M \leq \sqrt{N}$ and any map $\alpha_t^{y_t}:\sX \to \sX$. If Assumption \ref{BoundedAssumption} is satisfied and $|\mathcal{I}_t|=M$, then 
\begin{equation}
\| (\varphi,\pi_t^N) - (\varphi, \pi_t) \|_p \leq \frac{c_{t,p} \|\varphi\|_\infty}{\sqrt{N}}
\label{eqJJ0}
\end{equation}
for every $t=1, 2, ..., T$, any $\varphi \in B(\mathsf{X})$, any $p \ge 1$ and some constant $c_{t,p} < \infty$ independent of $N$.
\end{thm}

See Appendix \ref{apNudgingBound} for a proof.

Theorem \ref{ThmNudgingBound} is very general; it actually holds for any map $\alpha_t^{y_t}: \sX \to \sX$, i.e., not necessarily a nudging operator. We can also obtain error rates for specific choices of the nudging scheme. A simple, yet practically appealing, setup is the combination of batch and gradient nudging, as described in Sections \ref{secHowToChoose} and \ref{secHowToNudge}, respectively. 

\begin{assumption} \label{GradientBounded} 
The gradient of the likelihood is bounded. In particular, there are constants $G_t<\infty$ such that
\begin{align*}
\|\nabla_x g_t(x)\|_2 \leq G_t &< \infty
\end{align*}
for every $x \in \mathsf{X}$ and $t=1, 2, \ldots, T$.
\end{assumption}

\begin{lem}\label{LemGradientBound} 
Choose the number of nudged particles, $M >0$, and a sequence of step-sizes, $\gamma_t>0$, in such a way that $\sup_{1 \le t \le T} \gamma_t M \leq \sqrt{N}$ for some $T<0$. If Assumption \ref{GradientBounded} holds and $\varphi$ is a Lipschitz test function, then the error introduced by the batch gradient nudging step with $|\mathcal{I}_t|=M$ can be bounded as,
\begin{align*}
\left\| (\varphi, {\xi}_t^N) - (\varphi, \tilde{\xi}_t^N) \right\|_p \leq \frac{LG_t}{\sqrt{N}},
\end{align*}
where $L$ is the Lipschitz constant of $\varphi$, for every $t=1, \ldots, T$.
\end{lem}
See Appendix \ref{apLemGradientBound} for a proof. 

It is straightforward to apply Lemma \ref{LemGradientBound} to prove convergence of the NuPF with a batch gradient-nudging step. Specifically, we have the following result.

\begin{thm}\label{ThmGradientCorollary} 
Let $y_{1:T}$ be an arbitrary but fixed sequence of observations, with $T<\infty$, and choose a sequence of step sizes $\gamma_t>0$ and an integer $M$ such that $$\sup_{1\le t\le T} \gamma_t M \leq \sqrt{N}.$$ Let $\pi_t^N$ denote the filter approximation obtained with a NuPF with batch gradient nudging. If Assumptions \ref{BoundedAssumption} and \ref{GradientBounded} are satisfied and $|\mathcal{I}_t|=M$, then 
\begin{equation}
\| (\varphi,\pi_t^N) - (\varphi, \pi_t) \|_p \leq \frac{c_{t,p} \|\varphi\|_\infty}{\sqrt{N}}
\end{equation}
for every $t=1, 2, ..., T$, any bounded Lipschitz function $\varphi$, some constant $c_{t,p} < \infty$ independent of $N$ for any integer $p \geq 1$.
\end{thm}

The proof is straightforward (using the same argument as in the proof of Theorem \ref{ThmNudgingBound} combined with Lemma \ref{LemGradientBound}) and we omit it here. We note that Lemma \ref{LemGradientBound} provides a guideline for the choice of $M$ and $\gamma_t$. In particular, one can select $M = N^\beta$, where $0 < \beta < 1$, together with $\gamma_t \leq N^{\frac{1}{2} - \beta}$ in order to ensure that $\gamma_t M \leq \sqrt{N}$. Actually, it would be sufficient to set $\gamma_t \leq C N^{\frac{1}{2} - \beta}$ for some constant $C<\infty$ in order to keep the same error rate (albeit with a different constant in the numerator of the bound). Therefore,  Lemma \ref{LemGradientBound} provides a heuristic to balance the step size with the number of nudged particles\footnote{{Note that the step sizes may have to be kept small enough to ensure that $g_t(\bar  x_t^{(i)} + \gamma_t \nabla_x g_t(\bar x_t^{(i)}) ) \ge g_t(\bar x_t^{(i)})$, so that proper nudging, according to Definition \ref{defNudging}, is performed.}}. We can increase the number of nudged particles but in that case we need to shrink the step size accordingly, so as to keep $\gamma_t M \leq \sqrt{N}$. Similar results can be obtained using the gradient of the log-likelihood, $\log g_t$, if the $g_t$ comes from the exponential family of densities.

%
\subsection{Uniform convergence} \label{ssUniform}
Uniform convergence can be proved for the NuPF under the same standard assumptions as for the conventional BPF; see, e.g., \citet{del2001stability,del2004feynman}. The latter can be summarised as follows \citep{del2004feynman}: 
\begin{itemize}
\item[(i)] The likelihood function is bounded and bounded away from zero, i.e., $g_t \in B(\mathsf{X})$ and there is some constant $a>0$ such that $\inf_{t>0, x\in\mathsf{X}} g_t(x) \ge a$.
\item[(ii)] The kernel mixes sufficiently well, namely,  for any given integer $m$ there is a constant $0<\varepsilon<1$ such that
\begin{align*}
\inf_{t>0;(x,x')\in\mathsf{X}^2} \frac{
	\tau_{t+m|t}(A|x)
}{
	\tau_{t+m|t}(A|x')
} > \varepsilon
\end{align*}
for any Borel set A, where $\tau_{t+m|t}$ is the composition of the kernels $\tau_{t+m} \circ \tau_{t+m-1} \circ \cdots \circ \tau_t$.
\end{itemize}
When (i) and (ii) above hold, the sequence of optimal filters $\{ \pi_t \}_{t\ge 0}$ is stable and it can be proved that 
\begin{align*}
\sup_{t>0} \| (\varphi,\pi_t) - (\varphi,\pi_t^N) \|_p \le \frac{c_p}{\sqrt{N}}
\end{align*}
for any bounded function $\varphi\in B(\mathsf{X})$, where $c_p<\infty$ is constant with respect to $N$ and $t$ and $\pi_t^N$ is the particle approximation produced by either the NuPF (as in Theorem \ref{ThmNudgingBound} or, provided $\sup_{t>0} G_t < \infty$, as in Theorem \ref{ThmGradientCorollary}) or the BPF algorithms. We skip a formal proof as, again, it is straightforward combination of the standard argument by \citet{del2004feynman} (see also, e.g., \citet{Oreshkin11} and \citet{Crisan17}) with the same handling of the nudging operator as in the proofs of Theorem \ref{ThmNudgingBound} or Lemma \ref{LemGradientBound} .

%
\subsection{Nudging as a modified dynamical model} \label{ssModelling}
We have found in computer simulation experiments that the NuPF is consistently more robust to model errors than the conventional BPF. In order to obtain some analytical insight of this scenario, in this section we reinterpret the NuPF as a standard BPF for a modified, observation-driven dynamical model and discuss why this modified model can be expected to be a better fit for the given data than the original SSM. In this way, the NuPF can be seen as an automatic adaptation of the underlying model to the available data.

The dynamic models of interest in stochastic filtering can be defined by a prior measure $\tau_0$, the transition kernels $\tau_t$ and the likelihood functions $g_t(x)=g_t(y_t|x)$, for $t\ge 1$. In this section we write the latter as $g_t^{y_t}(x) = g_t(y_t|x)$, in order to emphasize that $g_t$ is parametrised by the observation $y_t$, and we also assume that every $g_t^{y_t}$ is a normalised pdf in $y_t$ for the sake of clarity. Hence, we can formally represent the SSM defined by \eqref{prioreq}, \eqref{transitionEq} and \eqref{observationEq} as $\cM_0=\{\tau_0,\tau_t,g_t^{y_t}\}$.


Now, let us assume $y_{1:T}$ to be fixed and construct the alternative {dynamical model} $\cM_1 = \{ \tau_0, \tilde \tau_t^{y_t}, g_t^{y_t} \}$, where
\begin{align}
\tilde \tau_t^{y_t}(\md x_t | x_{t-1} ) := &(1-\varepsilon_M)\tau_t(\md x_t|x_{t-1}) \, + \nonumber \\
& \varepsilon_M \int \delta_{\alpha_t^{y_t}(\bar{x}_t)} (\md x_t) \tau_t(\md \bar{x}_t | x_{t-1})
\label{eqAltSSM}
\end{align}
{is an observation-driven transition kernel}, $\varepsilon_M = \frac{M}{N}$ and the nudging operator $\alpha_t^{y_t}$ is a one-to-one map that depends on the (fixed) observation $y_t$. We note that the kernel $\tilde \tau_t^{y_t}$ jointly represents the Markov transition induced by the original kernel $\tau_t$ followed by an independent nudging transformation (namely, each particle is independently nudged with probability $\varepsilon_M$). As a consequence, the standard BPF for model $\cM_1$ coincides exactly with a NuPF for model $\cM_0$ with independent nudging and operator $\alpha_t^{y_t}$. Indeed, according to the definition of $\tilde \tau_t^{y_t}$ in \eqref{eqAltSSM}, generating a sample $\tilde x_t^{(i)}$ from $\tilde \tau_t^{y_t}(\md x_t | x_{t-1}^{(i)})$ is a three-step process where 
\begin{itemize}
\item we first draw $\bar x_t^{(i)}$ from $\tau_t(\md x_t|x_{t-1}^{(i)})$, 
\item then generate a sample $u_t^{(i)}$ from the uniform distribution $\cU(0,1)$, and 
\item if $u_t^{(i)} < \varepsilon_M$ then we set $\tilde x_t^{(i)} = \alpha_t^{y_t}(\bar x_t^{(i)})$, else we set $\tilde x_t^{(i)} = \bar x_t^{(i)}$.
\end{itemize}
After sampling, the importance weight for the BPF applied to model $\cM_1$ is $w_t^{(i)} \propto g_t^{y_t}(\tilde x_t^{(i)})$. This is exactly the same procedure as in the NuPF applied to the original SSM $\cM_0$ (see Algorithm \ref{Algnudged}).

Intuitively, one can expect that the observation-driven $\cM_1$ is a better fit for the data sequence $y_{1:T}$ than the original model $\cM_0$. Within the Bayesian methodology, a common approach to compare two competing probabilistic models ($\cM_0$ and $\cM_1$ in this case) for a given data set $y_{1:t}$ is to evaluate the so-called {\em model evidence} \citep{Bernardo94} for both $\cM_0$ and $\cM_1$. 
\begin{defn}
The evidence (or likelihood) of a probabilistic model $\cM$ for a given data set $y_{1:t}$ is the probability density of the data conditional on the model, that we denote as $\sfp(y_{1:t}|\cM)$.
\end{defn}
We say that $\cM_1$ is a better fit than $\cM_0$ for the data set $y_{1:t}$ when $\sfp(y_{1:t}|\cM_1) > \sfp(y_{1:t}|\cM_0)$. Since
\begin{equation}
\sfp(y_{1:t}|\cM_0) = \int\cdots\int \prod_{l=1}^t g_l(x_l)\tau_l(\md x_l | x_{l-1}) \tau_0(\md x_0), 
\nonumber
\end{equation}
and
\begin{equation}
\sfp(y_{1:t}|\cM_1) = \int\cdots\int \prod_{l=1}^t g_l(x_l)\tilde \tau_l^{y_l}(\md x_l | x_{l-1}) \tau_0(\md x_0), 
\nonumber
\end{equation}
the difference between the evidence of $\cM_0$ and the evidence of $\cM_1$ depends on the difference between the transition kernels $\tilde \tau_t$ and $\tilde \tau_t^{y_t}$.

We have empirically observed in several computer experiments that $\sfp(y_{1:t}|\cM_1) > \sfp(y_{1:t}|\cM_0)$ and we argue that the observation-driven kernel $\tilde \tau_t^{y_t}$ implicit to the NuPF is the reason why the latter filter is robust to modelling errors in the state equation, compared to standard PFs. This claim is supported by the numerical results in Sections \ref{ssLorenz63} and \ref{ssObjectTracking}, which show how the NuPF attains a significant better performance than the standard BPF, the auxiliary PF \cite{pitt1999filtering} or the extended Kalman filter \citep{Anderson79} in scenarios where the filters are built upon a transition kernel different from the one used to generate the actual observations.

While it is hard to show that $\sfp(y_{1:t}|\cM_1) > \sfp(y_{1:t}|\cM_0)$ for {\em every} NuPF, it is indeed possible to guarantee that the latter inequality holds for specific nudging schemes. An example is provided in Appendix \ref{apEvidence}, where we describe a certain nudging operator $\alpha_t^{y_t}$ and then proceed to prove that $\sfp(y_{1:t}|\cM_1) > \sfp(y_{1:t}|\cM_0)$, for that particular scheme, under some regularity conditions on the likelihoods and transition kernels.

\section{Computer simulations}\label{secNumericalSims}

In this section, we present the results of several computer experiments. In the first one, we address the tracking of a linear-Gaussian system. This is a very simple model which enables a clearcut comparison of the NuPF and other competing schemes, including a conventional PF with optimal importance function (which is intractable for all other examples) and a PF with nudging and proper importance weights. Then, we study three nonlinear tracking problems: 
\begin{itemize}
\item a stochastic Lorenz 63 model with misspecified parameters, 
\item a maneuvering target monitored by a network of sensors collecting nonlinear observations corrupted with heavy-tailed noise, 
\item and, finally,  a high-dimensional stochastic Lorenz 96 model\footnote{For the experiments involving Lorenz 96 model, simulation from the model is implemented in C++ and integrated into Matlab. The rest of the simulations are fully implemented in Matlab.}.
\end{itemize}

We have used gradient nudging in all experiments, with either $M \leq \sqrt{N}$ (deterministically, with batch nudging) or $\bE[M]\le \sqrt{N}$ (with independent nudging). {This ensures that the assumptions of Theorem \ref{ThmNudgingBound} hold. For simplicity, the gradient steps are computed with fixed step sizes, i.e., $\gamma_t=\gamma$ for all $t$.} For the object tracking experiment, we have used a large step-size, but this choice does not affect the convergence rate of the NuPF algorithm either.

%
\begin{figure*}[t]
\begin{center}
\includegraphics[scale=0.43]{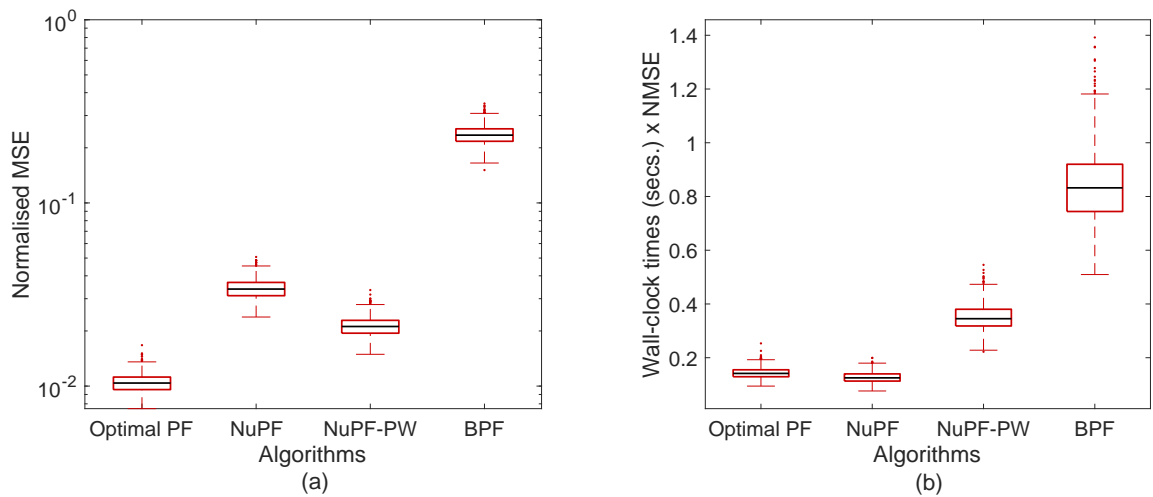}
\end{center}
\caption{(a) $\overline{\mbox{NMSE}}$ of the Optimal PF, NuPF-PW, NuPF, and BPF methods implemented for the high-dimensional linear-Gaussian SSM given in \eqref{eq:priorLGSSM}--\eqref{eq:likelihoodLGSSM}. The box-plots are constructed from 1,000 independent Monte Carlo runs. It can be seen that the $\overline{\mbox{NMSE}}$ of the NuPF is comparable to the error of the Optimal PF and the NuPF-PW methods. (b) Runtimes$\times$$\overline{\mbox{NMSE}}$s of all methods. This experiment shows that, in addition to the fact that the NuPF attains a comparable estimation performance, which can be seen in (a), it has a computational cost similar to the plain BPF. The figure demonstrates that the NuPF has a comparable performance to the optimal PF for this model.}
\label{FigOptimalCompare}
\end{figure*}

\subsection{A high-dimensional, inhomogeneous Linear-Gaussian state-space model}\label{secKalmanNudging}

In this experiment we compare different PFs implemented to track a high-dimensional linear Gaussian SSM. In particular, the model under consideration is
\begin{align}
x_0 &\sim \NPDF(0,I_{d_x}), \label{eq:priorLGSSM}\\
x_t|x_{t-1} &\sim \NPDF(x_{t-1},Q),\label{eq:transLGSSM}\\
y_t|x_t &\sim \NPDF(C_t x_t, R), \label{eq:likelihoodLGSSM}
\end{align}
where $\{x_t\}_{t\geq 0}$ are hidden states, $\{y_t\}_{t\geq 1}$ are observations, and $Q$ and $R$ are the process and the observation noise covariance matrices, respectively. The latter are diagonal matrices, namely $Q = q I_{d_x}$ and $R = I_{d_y}$, where $q = 0.1$, $d_x = 100$ and $d_y = 20$. The sequence $\{C_t\}_{t\geq 1}$ defines a time-varying observation model. The elements of this sequence are chosen as random binary matrices, i.e., $C_t \in \{0,1\}^{d_y \times d_x}$ where each entry is generated as an independent Bernoulli random variable with $p = 0.5$. Once generated, they are fixed and fed into all algorithms we describe below for each independent Monte Carlo run.

We compare the NuPF with three alternative PFs. The first method we implement is the PF with the optimal proposal pdf $p(x_t|x_{t-1},y_t) \propto g_t(y_t|x_t) \tau_t(x_t|x_{t-1})$, abbreviated as Optimal PF. The pdf $p(x_t|x_{t-1},y_t)$ leads to an analytically tractable Gaussian density for the model \eqref{eq:priorLGSSM}--\eqref{eq:likelihoodLGSSM} \citep{doucet2000sequential} but not in the nonlinear tracking examples below. Note, however, that at each time step, the mean and covariance matrix of this proposal have to be explicitly evaluated in order to compute the importance weights. 

The second filter is a nudged PF with proper importance weights (NuPF-PW). In this case, we treat the generation of the nudged particles as a proposal function to be accounted for during the weighting step. To be specific, the proposal distribution resulting from the NuPF has the form
\begin{align}\label{eq:mixtureNuPF}
\tilde{\tau}_t(\mbox{d}x_t | x_{t-1}) = (1-\epsilon_N) \tau_t(\mbox{d}x_t|x_{t-1}) + \epsilon_N \bar{\tau}_t(\mbox{d}x_t|x_{t-1}),
\end{align}
where $\epsilon_N = \frac{1}{\sqrt{N}}$ and
\begin{align*}
\bar{\tau}_t(\mbox{d}x_t|x_{t-1}) = \int \delta_{\alpha_t^{y_t}(\bar{x}_t)} (\mbox{d}x_t) \tau_t(\mbox{d}\bar{x}_t|x_{t-1}).
\end{align*}
The latter conditional distribution admits an explicit representation as a Gaussian for model \eqref{eq:priorLGSSM}-\eqref{eq:likelihoodLGSSM} when the $\alpha_t$ operator is designed as a gradient step, but this approach is intractable for the examples in Section \ref{ssLorenz63} and Section \ref{ssLorenz96}. Note that $\tilde \tau_t$ is a mixture of two time-varying Gaussians and this fact adds to the cost of the sampling and weighting steps. Specifically, computing weights for the NuPF-PW is significantly more costly, compared to the BPF or the NuPF, because the mixture \eqref{eq:mixtureNuPF} has to be evaluated together with the likelihood and the transition pdf. 

The third tracking algorithm implemented for model \eqref{eq:priorLGSSM}--\eqref{eq:likelihoodLGSSM} is the conventional BPF.

For all filters, we have set the number of particles as\footnote{{When $N$ is increased the results are similar for the NuPF, the optimal PF and the NuPF-PW larger number particles, as they already perform close to optimally for $N=100$, and only the BPF improves significantly.}} $N = 100$ . In order to implement the NuPF and NuPF-PW schemes, we have selected the step size $\gamma = 2 \times 10^{-2}$. We have run 1,000 independent Monte Carlo runs for this experiment. To evaluate different methods, we have computed the empirical normalised mean squared errors (NMSEs). {Specifically, the NMSE for the $j$-th simulation is 
\begin{align}\label{eqNMSE}
\overline{\mbox{NMSE}}(j) =  \frac{\sum_{t=1}^{t_f} \| \bar{x}_t - \hat x_t(j) \|_2^2}{\sum_{t=1}^{t_f} \| x_t \|_2^2},
\end{align}
where $\bar{x}_t = \mathbb{E}[x_t|y_{1:t}]$ is the exact posterior mean of the state $x_t$ conditioned on the observations up to time $t$ and $\hat x_t(j)$ is the estimate of the state vector in the $j$-th simulation run. Therefore, the notation $\overline{\mbox{NMSE}}$ implies the normalised mean squared error is computed with respect to $\bar{x}_t$. In the figures, we usually plot the mean and the standard deviation of the sample of errors, $\overline{\mbox{NMSE}}(1), \ldots, \overline{\mbox{NMSE}}(1000)$.}

The results  are shown in Fig. \ref{FigOptimalCompare}. In particular, in Fig.~\ref{FigOptimalCompare}(a), we observe that the $\overline{\mbox{NMSE}}$ performance of the NuPF compared to the optimal PF and NuPF-PW (which is similar to a classical PF with nudging) is comparable. However, Fig.~\ref{FigOptimalCompare}(b) reveals that the NuPF is significantly less demanding compared to the optimal PF and the NuPF-PW method. Indeed, the run-times of the NuPF are almost identical to the those of the plain BPF. As a result, the plot of the $\overline{\mbox{NMSE}}$s multiplied by the running times displayed in Fig.~\ref{FigOptimalCompare}(b) reveals that the proposed algorithm is as favorable as the optimal PF, which can be implemented for this model, but not for general models unlike the NuPF.

\begin{figure*}[t]
\begin{center}
\includegraphics[scale=0.62]{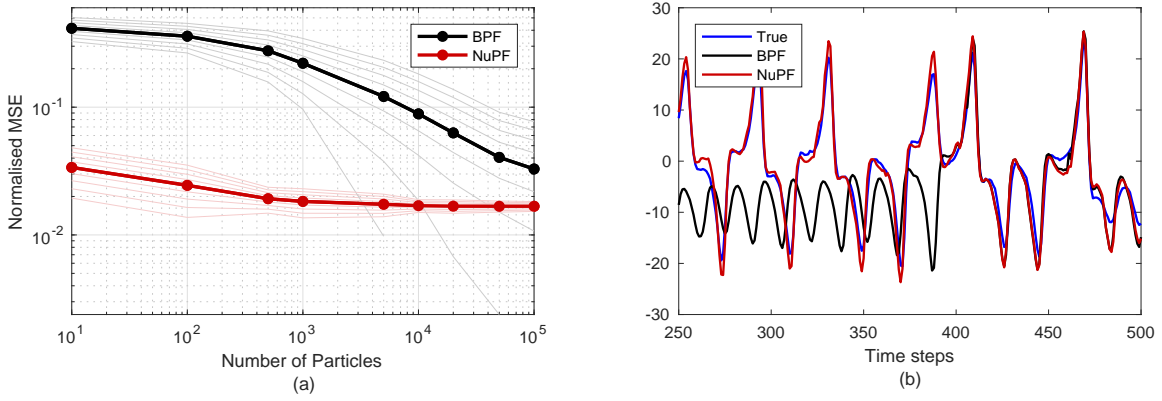}
\end{center}
\caption{(a) NMSE results of the BPF and NuPF algorithms for a misspecified Lorenz 63 system. The results have been obtained from 1,000 independent Monte Carlo runs for each $N \in \{10,100,500,1K,5K,10K,20K,50K,100K\}$. The {light-coloured lines indicate the area containing up to one standard deviation from the empirical mean}. The misspecified parameter is $\mathsf{b}_\epsilon = \mathsf{b} + \epsilon$, where $\mathsf{b} = 8/3$ and $\epsilon = 0.75$. (b) A sample path of the true state variable $x_{2,t}$ and its estimates in a run with $N = 500$ particles.}\label{FigLorenz63}
\end{figure*}
%
\subsection{Stochastic Lorenz 63 model with misspecified parameters} \label{ssLorenz63}

In this experiment, we demonstrate the performance of the NuPF {when tracking a} misspecified stochastic Lorenz 63 model. The dynamics of the system is described by a stochastic differential equation (SDE) in three dimensions,
\begin{align*}
\mbox{d} x_1 &= - \mathsf{a}(x_1 - x_2)\md s + \mbox{d} w_1, \\
\mbox{d} x_2 &= \left( \mathsf{r}x_1 - x_2 - x_1 x_3 \right) \md s + \mbox{d} w_2, \\
\mbox{d} x_3 &= \left( x_1 x_2 - \mathsf{b}x_3 \right) \md s + \mbox{d} w_3,
\end{align*}
where $s$ denotes continuous time, $\{w_i(s)\}_{s\in(0,\infty)}$ for $i = 1,2,3$ are 1-dimensional independent Wiener processes and $\mathsf{a},\mathsf{r},\mathsf{b} \in \bR$ are fixed model parameters. We discretise the model using the Euler-Maruyama scheme with integration step $\mathsf{T} > 0$ and obtain the system of difference equations
\begin{eqnarray}
x_{1,t} &=& x_{1,t-1} - \mathsf{T} \mathsf{a} (x_{1,t-1} - x_{2,t-1}) \nonumber \\
&& + \sqrt{\mathsf{T}} u_{1,t}, \label{eqEulerM} \\
x_{2,t} &=& x_{2,t-1} + \mathsf{T} (\mathsf{r} x_{1,t-1} -
x_{2,t-1} - x_{1,t-1} x_{3,t-1}) \nonumber \\
&& + \sqrt{\mathsf{T}} u_{2,t}, \nonumber \\
x_{3,t} &=& x_{3,t-1} + \mathsf{T} (x_{1,t-1} x_{2,t-1} - \mathsf{b}x_{3,t-1}) + \sqrt{\mathsf{T}} u_{3,t}, \nonumber
\end{eqnarray}
where $\{u_{i,t}\}_{t\in\bN}$, $i = 1,2,3$ are i.i.d. Gaussian random variables with zero mean and unit variance. We assume that we can only observe the variable $x_{1,t}$, {contaminated by additive noise, every $t_s>1$ discrete time steps}. To be specific, we collect the sequence of observations
\begin{align*}
y_n = k_o x_{1,n t_s} + v_n, \quad n = 1, 2, ...,
\end{align*}
where $\{v_n\}_{n\in\bN}$ is a sequence of i.i.d. Gaussian random variables with zero mean and unit variance and the scale parameter $k_o = 0.8$ is assumed known. 

In order to simulate both the state signal and the synthetic observations from this model, we choose the so-called standard parameter values 
\begin{align*}
(\mathsf{a,r,b}) = \left(10,28,\frac{8}{3}\right),
\end{align*}
which make the system dynamics chaotic. The initial condition is set as 
\begin{align*}
x_0 = [-5.91652, -5.52332, 24.5723]^\top.
\end{align*}
{The latter value has been chosen from a deterministic trajectory of the system (i.e., with no state noise) with the same parameter set  $(\mathsf{a,r,b}) = \left(10,28,\frac{8}{3}\right)$ to ensure that the model is started at a sensible point.} We assume that the system is observed every $t_s = 40$ discrete time steps and for each simulation we simulate the system for $t=0, 1, \ldots, t_f$, with $t_f = 20,000$. Since $t_s = 40$, we have a sequence of $\frac{t_f}{t_s}=500$ observations overall. 

Let us note here that the Markov kernel which takes the state from time $n-1$ to time $n$ (i.e., from the time of one observation to the time of the next observation) is straightforward to simulate using the Euler-Maruyama scheme \eqref{eqEulerM}, however the associated transition probability density cannot be evaluated because it involves the mapping of both the state and a sequence of $t_s$ noise samples through a composition of nonlinear functions. This precludes the use of importance sampling schemes that require the evaluation of this density when computing the weights.

We run the BPF and NuPF algorithms for the model described above, except that the parameter $\mathsf{b}$ is replaced by $\mathsf{b}_\epsilon = \mathsf{b}+\epsilon$, with $\epsilon = 0.75$ (hence $\mathsf{b}_\epsilon \approx 3.417$ versus $\mathsf{b} \approx 2.667$ for the actual system). As the system underlying dynamics is chaotic, this mismatch affects the predictability of the system significantly.

\begin{figure*}
\begin{center}
\includegraphics[scale=0.5]{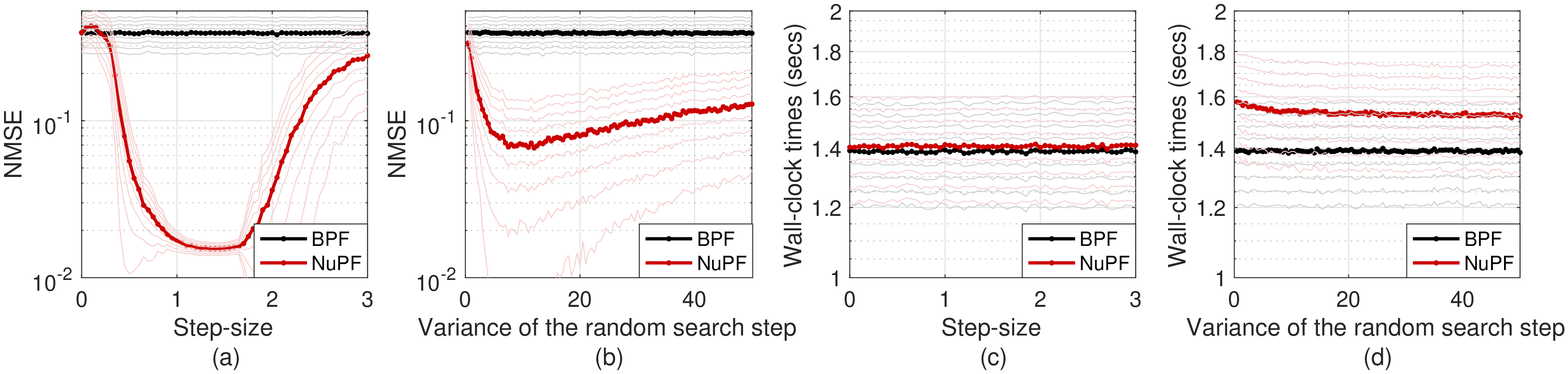}
\end{center}
\caption{A comparison of gradient nudging and random search nudging for a variety of parameter settings. From (a), it can be seen that gradient nudging is robust within a large interval for $\gamma$. From (b), one can see that the same is true for random search nudging with the covariance of the form $C = \sigma^2 I$ for a wide range of $\sigma^2$. From (c)--(d), it can be seen that while gradient nudging causes negligible computational overhead, random search nudging is more demanding in terms of computation time and this behaviour is expected to be more apparent in higher dimensional spaces. Comparing (a)--(b), it can also be seen that gradient nudging attains lower error rates in general. {The lighter-coloured lines indicate the area containing up to one standard deviation from the empirical means in each plot.}}\label{FigRandSearch}
\end{figure*}

We have implemented the NuPF with independent gradient nudging. Each particle is nudged with probability $\frac{1}{\sqrt{N}}$, where $N$ is the number of particles (hence $\bE[M]=\sqrt{N}$), and the size of the gradient steps is set to $\gamma = 0.75$ (see Algorithm \ref{gradientNudging}). 

{As a figure of merit,} we evaluate the NMSE for the 3-dimensional state vector, averaged over 1,000 independent Monte Carlo simulations. For this example (as well as in the rest of this section), it is not possible to compute the exact posterior mean of the state variables. Therefore, the NMSE values are computed with respect to the ground truth, i.e.,
\begin{align}\label{eqNMSEgroundtruth}
{\mbox{{NMSE}}}(j) =  \frac{\sum_{t=1}^{t_f} \| {x}_t - \hat x_t(j) \|_2^2}{\sum_{t=1}^{t_f} \| x_t \|_2^2},
\end{align}
where $(x_t)_{t\geq 1}$ is the ground truth signal.

Fig.~\ref{FigLorenz63} (a) displays the ${\mbox{{NMSE}}}$, attained for varying number of particles $N$, for the standard BPF and the NuPF. It is seen that the NuPF outperforms the BPF for the whole range of values of $N$ in the experiment, both in terms of the mean and the standard deviation of the errors, although the NMSE values become closer for larger $N$. The plot on the right displays the values of $x_{2,t}$ and its estimates for a typical simulation. In general, the experiment shows that the NuPF can track the actual system using the misspecified model and a small number of particles, whereas the BPF requires a higher computational effort to attain a similar performance.

As a final experiment with this model, we have tested the robustness of the algorithms with respect to the choice of parameters in the nudging step. In particular, we have tested the NuPF with independent gradient nudging for a wide range of step-sizes $\gamma$. Also, we have tested the NuPF with random search nudging using a wide range of covariances of the form $C = \sigma^2 I$ by varying $\sigma^2$.

{The results can be seen in Fig.~\ref{FigRandSearch}. This figure shows that the algorithm is robust to the choice of parameters for a range of step-sizes and variances of the random search step. As expected, random search nudging takes longer running time compared to gradient steps. This difference in run-times is expected to be larger in higher-dimensional models since random search is expected to be harder in such scenarios.}

%
\begin{figure*}[t]
\begin{center}
\includegraphics[scale=0.44]{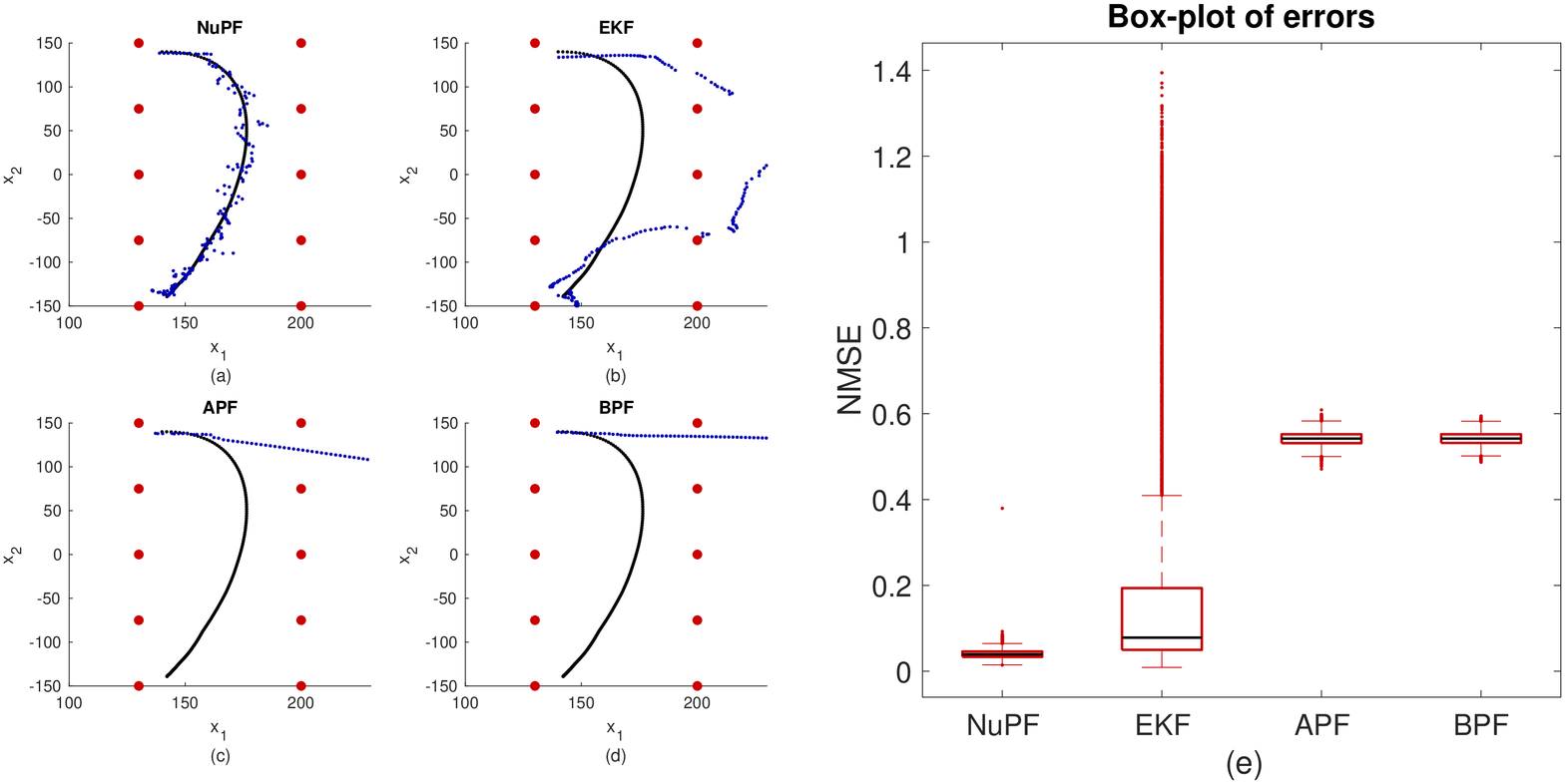}
\caption{Plots (a)--(d): A typical simulation run for the BPF, APF, EKF and NuPF algorithms using $N=500$ particles. The black dots denote the real trajectory of the object, the red dots are sensors and the blue dots are position estimates as provided by the filters. Plot (e): Box-plot of the errors $\mbox{NMSE}(1), \ldots, \mbox{NMSE}(10,000)$ obtained for the set of independent simulation runs. The NuPF achieves a low NMSE with a low variance whereas the EKF exhibits a large variance.}
\label{FigObjTracking}
\end{center}
\end{figure*}

\subsection{Object tracking with a misspecified model} \label{ssObjectTracking}

In this experiment, we consider a tracking scenario where a target is observed through sensors collecting radio signal strength (RSS) measurements contaminated with additive heavy-tailed noise. The target dynamics are described by the model,
\begin{align*}
x_t &= A x_{t-1} + B L (x_{t-1} - x_{{\bullet}}) + u_t,
\end{align*}
where $x_t\in\bR^4$ denotes the target state, consisting of its position $r_t \in \bR^2$ and its velocity, $v_t \in \bR^2$, hence $x_t = \left[ \begin{array}{c} r_t\\v_t\\ \end{array} \right]\in\bR^4$. The vector $x_{{\bullet}}$ is a deterministic, pre-chosen state to be attained by the target. Each element in the sequence $\{u_t\}_{t\in\bN}$ is a zero-mean Gaussian random vector with covariance matrix $Q$. The parameters $A,B,Q$ are selected as
\begin{align*}
A = \begin{bmatrix}
    I_2 & \kappa I_2\\
    0   & 0.99 I_2
\end{bmatrix}, \quad
B = \begin{bmatrix}
    0 & I_2
\end{bmatrix}^\top,
\end{align*}
and
\begin{align*}
Q = \begin{bmatrix}
    \frac{\kappa^3}{3} I_2 & \frac{\kappa^2}{2} I_2 \\
    \frac{\kappa^2}{2} I_2 & \kappa I_2
\end{bmatrix},
\end{align*}
where $I_2$ is the $2\times 2$ identity matrix and $\kappa = 0.04$. The policy matrix $L \in \bR^{2\times4}$ determines the trajectory of the target from an initial position $x_0 = [140,140,50,0]^\top$ to a final state $x_T = [140,-140,0,0]^\top$ and it is computed by solving a Riccati equation (see \citet{BertsekasOptCont1} for details), which yields
\begin{align*}
L = \left[
	\begin{array}{cccc}
	-0.0134& 0 & -0.0381 & 0\\
	0 & -0.0134 & 0 & -0.0381\\
	\end{array}
\right].
\end{align*}
This policy results in a highly maneuvering trajectory. In order to design the NuPF, however, we assume the simpler dynamical model 
\begin{align*}
x_t &= A x_{t-1} + u_t,
\end{align*}
hence there is a considerable model mismatch.

The observations are nonlinear and coming from $10$ sensors placed in the region where the target moves. The measurement collected at the $i$-th sensor, time $t$, is modelled as 
\begin{align*}
y_{t,i} = 10 \log_{10}\left(\frac{P_0}{\|r_t - s_i\|^2} + \eta\right) + w_{t,i},
\end{align*}
where $r_t \in \bR^2$ is the location vector of the target, $s_i$ is the position of the $i$th sensor and $w_{t,i} \sim \mathcal{T}(0,1,\nu)$ is an independent t-distributed random variable for each $i = 1,\ldots,10$. Intuitively, the closer the parameter $\nu$ to $1$, the more explosive the observations become. In particular, we set $\nu = 1.01$ to make the observations explosive and heavy-tailed. As for the sensor parameters, we set the transmitted RSS as $P_0 = 1$ and the sensitivity parameter as $\eta = 10^{-9}$. The latter yields a soft lower bound of $-90$~decibels (dB) for the RSS measurements.

We have implemented the NuPF with batch gradient nudging, with a large-step size $\gamma = 5.5$ and $M = \lfloor \sqrt{N} \rfloor$. Since the observations depend on the position vector $r_t$ only, an additional model-specific nudging step is needed for the velocity vector $v_t$. In particular, after nudging the $r_t^{(i)} = [ x^{(i)}_{1,t}, x^{(i)}_{2,t}]^\top$, we update the velocity variables as
\begin{align*}
v_t^{(i)} = \frac{1}{\kappa} (r^{(i)}_{t} - r^{(i)}_{t-1}), \quad \mbox{where} \quad v_t^{(i)} =   [ x^{(i)}_{3,t}, x^{(i)}_{4,t}]^\top,
\end{align*}
where $\kappa= 0.04$ as defined for the model. The motivation for this additional transformation comes from the physical relationship between position and velocity. We note, however, that the NuPF also works without nudging the velocities.

We have run $10,000$ Monte Carlo runs with $N = 500$ particles in the auxiliary particle filter (APF) \citep{pitt1999filtering,johansen2008note,douc2009optimality}, the BPF \citep{gordon1993novel} and the NuPF. We have also implemented the extended Kalman filter (EKF), which uses the gradient of the observation model.
\begin{figure*}[t]
\begin{center}
\includegraphics[scale=0.73]{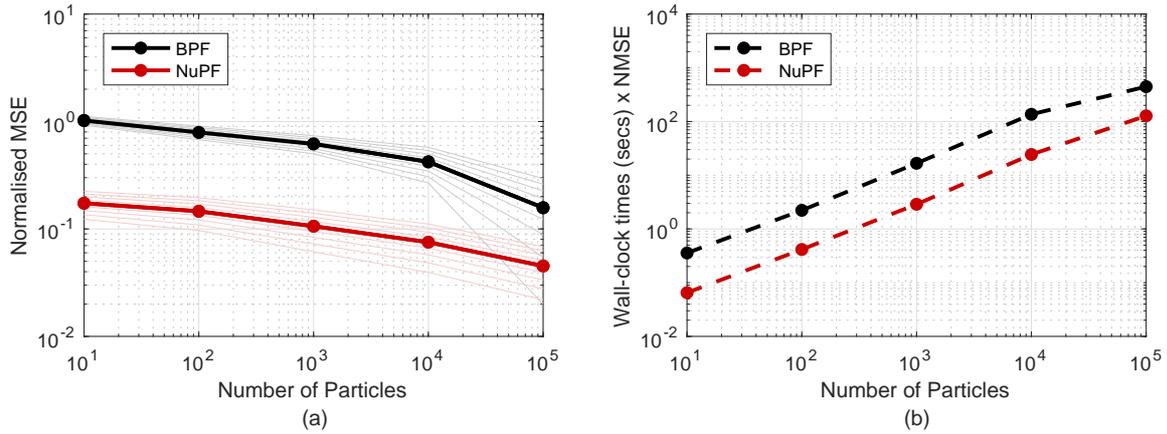}
\end{center}
\caption{Comparison of the NuPF and the BPF for the stochastic Lorenz 96 system with model dimension $d = 40$. The results have been averaged over a set of $1,024$ independent Monte Carlo runs. Plot (a): evolution of the NMSE as the number of particles $N$ is increased. {The light-coloured lines indicate the area containing up to one standard deviation from the empirical mean.} Plot (b): Run-times$\times$NMSE for the BPF and the NuPF in the same set of simulations. Since the increase in computational cost of the NuPF, compared to the BPF, is negligible, it is seen from plot (b) that the NuPF performs better when errors and run-times are considered jointly.}
\label{figBPFNPFtimeComparisond40}
\end{figure*}

Fig.~\ref{FigObjTracking} shows a typical simulation run with each one of the four algorithms (on the left side, plots (a)--(d)) and a box-plot of the NMSEs obtained for the 10,000 simulations (on the right, plot (e)). Plots (a)--(d) show that, while the EKF also uses the gradient of the observation model, it fails to handle the heavy-tailed noise, as it relies on Gaussian approximations. The BPF and the APF collapse due to the model mismatch in the state equation. Plot (d) shows that the NMSE of the NuPF is just slightly smaller in the mean than the NMSE of the EKF, but much more stable.

%
\subsection{High-dimensional stochastic Lorenz 96 model} \label{ssLorenz96}

In this computer experiment, we compare the NuPF with the ensemble Kalman filter (EnKF) for the tracking of a stochastic Lorenz 96 system. The latter is described by the set of stochastic differential equations (SDEs)
\begin{eqnarray}
\mbox{d} x_i &=& \left[ (x_{i+1} - x_{i-2}) x_{i-1} - x_i + F \right] \md s + \mbox{d} w_i, \nonumber\\
&& i = 1, \ldots, d, \nonumber
\end{eqnarray}
where $s$ denotes continuous time, $\{ w_i(s) \}_{s \in (0,\infty)}$, $1 \le i \le d$, are independent Wiener processes, $d$ is the system dimension and the forcing parameter is set to $F = 8$, which ensures a chaotic regime. The model is assumed to have a {circular spatial structure}, so that $x_{-1} = x_{d-1}$, $x_0 = x_d$, and $x_{d+1} = x_1$. {Note that each $x_i$, $i=1, ..., d$, denotes a time-varying state associated to a different space location.} In order to simulate data from this model, we apply the Euler-Maruyama discretisation scheme and obtain the difference equations,
\begin{eqnarray}
x_{i,t} &=& x_{i,t-1} + \mathsf{T} \left[ (x_{i+1,t-1} - x_{i-2,t-1}) x_{i-1,t-1} \right. \nonumber \\
&& \left. - x_{i,t-1} + F \right] + \sqrt{\mathsf{T}} u_{i,t}, \nonumber 
\end{eqnarray}
where $u_{i,t}$ are zero-mean, unit-variance Gaussian random variables. {We initialise this system by generating a vector from the uniform distribution on $(0,1)^d$ and running the system for a small number of iterations and set $x_0$ as the output of this short run.}

\begin{figure*}
\begin{center}
\includegraphics[scale=0.73]{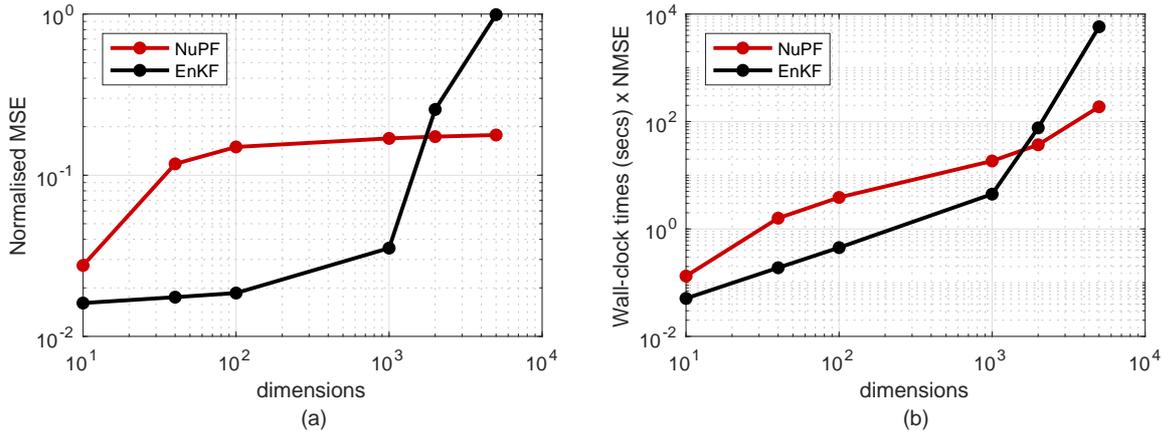}
\end{center}
\caption{Comparison of the NuPF with the EnKF for the stochastic Lorenz 96 model with increasing dimension $d$ and fixed number of particles $N = 500$ (this is the same as the number of ensemble members in the EnKF). We have run 1,000 independent Monte Carlo trials for this experiment. Plot (a): NMSE versus dimension $d$. The EnKF attains a smaller error for lower dimensions but then it explodes for $d > 10^3$, while the NuPF remains stable. Plot (b): Running-times$\times$NMSE plot for the same set of simulations. {It can be seen that the overall performance of the NuPF is better beyond 1K dimensions compared to the EnKF.}}
\label{figEnKFcomparison}
\end{figure*}

We assume that the system is only partially observed. In particular, half of the state variables are observed, in Gaussian noise, every $t_s = 10$ time steps, namely,
$$
y_{j,n} = x_{2j-1,nt_s} + u_{j,n},
$$
where $\quad n = 1, 2, \ldots$, $j = 1, 2, \ldots, \lfloor d/2 \rfloor$, and $u_{j,n}$ is a normal random variable with zero mean and unit variance. The same as in the stochastic Lorenz 63 example of Section \ref{ssLorenz63}, the transition pdf that takes the state from time $(n-1)t_s$ to time $nt_s$ is simple to simulate but hard to evaluate, since it involves mapping a sequence of noise variables through a composition of nonlinearities.

In all the simulations for this system we run the NuPF with batch gradient nudging (with $M = \lfloor \sqrt{N} \rfloor$ nudged particles and step-size $\gamma = 0.075$). In the first computer experiment, we fixed the dimension $d = 40$ and run the BPF and the NuPF with increasing number of particles. The results can be seen in Fig.~\ref{figBPFNPFtimeComparisond40}, which shows how the NuPF performs better than the BPF in terms of NMSE (plot (a)). {Since the run-times of both algorithms are nearly identical, it can be seen that, when considered jointly with NMSEs, the NuPF attains a significantly better performance (plot (b)).}

In a second computer experiment, we compared the NuPF with the EnKF. Fig.~\ref{figEnKFcomparison}(a) shows how the NMSE of the two algorithms grows as the model dimension $d$ increases and the number of particles $N$ is kept fixed. In particular, the EnKF attains a better performance for smaller dimensions (up to $d=10^3$), however its NMSE blows up for $d > 10^3$ while the performance of the NuPF remains stable. The running time of the EnKF was also higher than the running time of the NuPF in the range of higher dimensions ($d\ge 10^3$).

\subsection{Assessment of bias}

In this section, we numerically quantify the bias of the proposed algorithm on a low-dimensional linear-Gaussian state-space model. To assess the bias, we compute the marginal likelihood estimates given by the BPF and the NuPF. {The reason for this choice is that the BPF is known to yield unbiased estimates of the marginal likelihood\footnote{Note that the estimates of integrals $(\varphi,\pi_t)$ computed using the self-normalised importance sampling approximations (i.e., $(\varphi,\pi_t^N) \approx (\varphi,\pi_t)$) produced by the BPF and the NuPF methods are biased and the bias vanishes with the same rate for both algorithms as a result of Theorem~\ref{ThmNudgingBound}. The same is true for the approximate predictive measures $\xi_t^N$.} \citep{del2004feynman}. The NuPF leads to biased (typically overestimated) marginal likelihood estimates, hence it is of interest to compare them with those of the BPF.} To this end, we choose a simple linear-Gaussian state space model for which the marginal likelihood can be exactly computed as a byproduct of the Kalman filter. We then compare this exact marginal likelihood to the estimates given by the BPF and the NuPF.

Particularly, we define the state-space model,
\begin{align}
x_0 &\sim \NPDF(x_0;\mu_0,P_0),\\
x_t | x_{t-1} &\sim \NPDF(x_t; x_{t-1},Q), \label{eq:BiasSSMtrans} \\
y_t | x_t &\sim \NPDF(y_t; C_t x_t,R),\label{eq:BiasSSMobs}
\end{align}
where $(C_t)_{t\geq 0} \in [0,1]^{1 \times 2}$ is a sequence of observation matrices {where each entry is generated as a realisation of a Bernoulli random variable with $p = 0.5$}, $\mu_0$ is a zero vector, and $x_t \in \bR^2$ and $y_t \in \bR$. {The state variables are cross-correlated, namely,}
\begin{align*}
Q = \begin{bmatrix}
    2.7 & -0.48 \\
    -0.48 & 2.05
\end{bmatrix},
\end{align*}
and $R = 1$. We have chosen the prior covariance as $P_0 = I_{d_x}$. We have simulated the system for $T=100$ time steps. Given a fixed observation sequence $y_{1:T}$, the marginal likelihood for the system given in Eqs.~\eqref{eq:BiasSSMtrans}-\eqref{eq:BiasSSMobs} is
\begin{align*}
Z^\star = \mathsf{p}(y_{1:T}),
\end{align*}
which can be exactly computed via the Kalman filter.
\begin{figure*}[t]
\begin{center}
\includegraphics[scale=0.43]{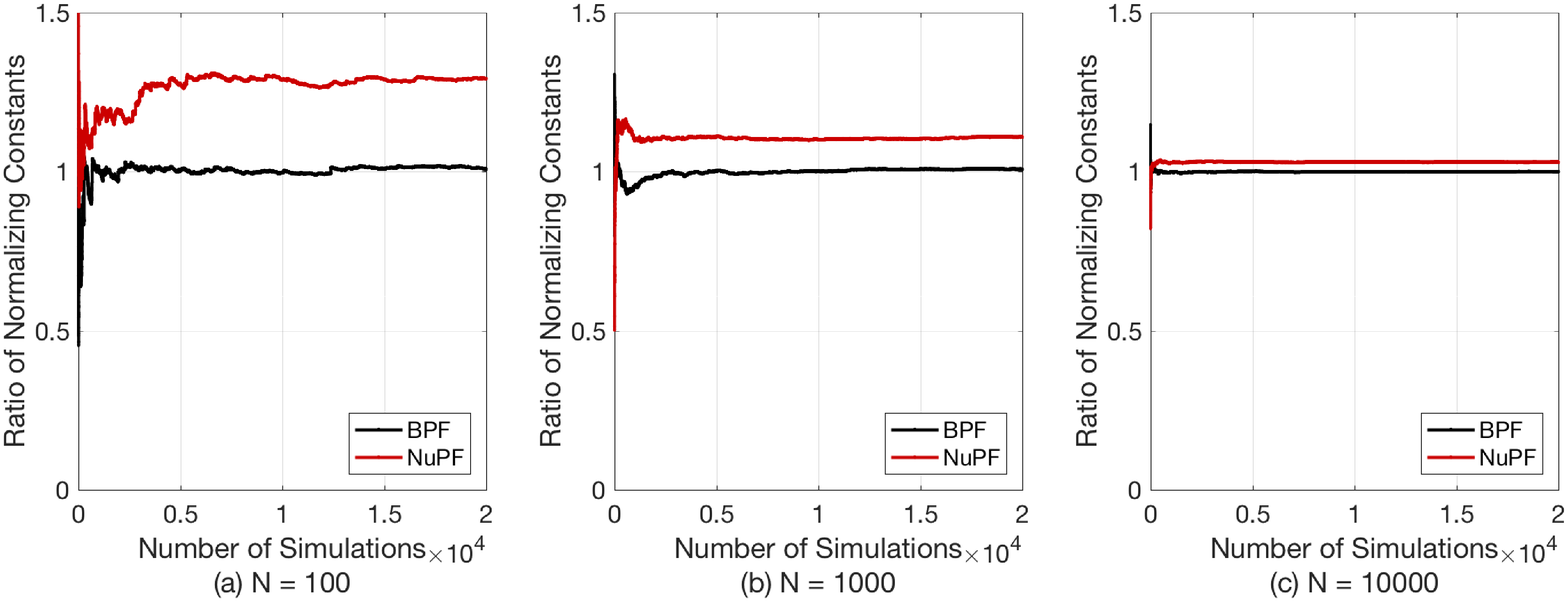}
\end{center}
\caption{{Evolution of the running averages $\bar{Z}^N_{\textnormal{BPF}}/Z^\star$ (black) and $\bar{Z}^N_{\textnormal{NuPF}}/Z^\star$ (red) for $K = 1, \ldots, 20,000$ independent simulations with $N = 100$, $N = 1,000$ and $N = 10,000$ particles for both filters. The ratio $\bar{Z}^N_{\textnormal{BPF}}/Z^\star$ for the BPF is unbiased \citep{del2004feynman} and hence converges to $1$. The ratio $\bar{Z}^N_{\textnormal{NuPF}}/Z^\star$ for the NuPF converges to $1+\epsilon$, with $\epsilon>0$ becoming smaller as $N$ increases, showing that the estimator $Z^N_{\textnormal{NuPF}}$ is biased (yet asymptotically unbiased with $N\rw\infty$; see Theorem \ref{ThmNudgingBound}).}}
\label{figBias}
\end{figure*}

We denote the estimate of $Z^\star$ given by the BPF and the NuPF as $Z^N_{\textnormal{BPF}}$ and $Z^N_{\textnormal{NuPF}}$, respectively. It is well-known that the BPF estimator is unbiased \citep{del2004feynman},
\begin{align}\label{eq:UnbiasednessProp}
\bE[Z^N_{\textnormal{BPF}}] = Z^\star,
\end{align}
where $\bE[\cdot]$ denotes the expectation with respect to the randomness of the particles. Numerically, this suggests that as one runs identical, independent Monte Carlo simulations to obtain $\{Z^{N,k}_{\textnormal{BPF}}\}_{k=1}^K$ and compute the average
\begin{align}\label{eq:BPFAverageBias}
\bar{Z}^{N}_{\textnormal{BPF}} = \frac{1}{K} \sum_{k=1}^K Z^{N,k}_{\textnormal{BPF}},
\end{align}
then it follows from the unbiasedness property \eqref{eq:UnbiasednessProp} that the ratio of the average in \eqref{eq:BPFAverageBias} and the true value $Z^\star$ should satisfy
\begin{align*}
\frac{\bar{Z}^{N}_{\textnormal{BPF}}}{Z^\star} \to 1 \quad \textnormal{as} \quad K \to \infty.
\end{align*}
Since the marginal likelihood estimates provided by the NuPF are not unbiased for the original SSM ({and tend to attain higher values}), if we define
\begin{align*}
\bar{Z}^{N}_{\textnormal{NuPF}} = \frac{1}{K} \sum_{k=1}^K Z^{N,k}_{\textnormal{NuPF}},
\label{eqRunningMeanNuPF}
\end{align*}
then as $K\to\infty$, we should see
\begin{align*}
\frac{\bar{Z}^{N}_{\textnormal{NuPF}}}{Z^\star} \to 1+\epsilon \quad \textnormal{as} \quad K \to \infty,
\end{align*}
for some $\epsilon > 0$.

{We have conducted an experiment aimed at quantifying the bias $\epsilon>0$ above. In particular, we have run 20,000 independent simulations for the BPF and the NuPF with $N=100$, $N=1,000$ and $N=10,000$. For each value of $N$, we have computed running empirical means as in \eqref{eq:BPFAverageBias} and \eqref{eqRunningMeanNuPF} for $K=1, \ldots, 20,000$. The variance of $\bar Z_{\textnormal{BPF}}^N$ increases with $T$, hence the estimators for small $K$ display a relatively large variance and we need $K>>1$ to clearly observe the bias. The NuPF filter performs independent gradient nudging with step size $\gamma = 0.1$}. 

{The results of the experiment are displayed in Figure \ref{figBias}, which shows how, as expected, the NuPF overestimates $Z^\star$. We can also see how the bias becomes smaller as $N$ increases (because only and average of $\sqrt{N}$ particles are nudged per time step).}

\section{Experimental results on model inference}\label{secExperiments}
\begin{figure*}[t]
\begin{center}
\includegraphics[scale=0.43]{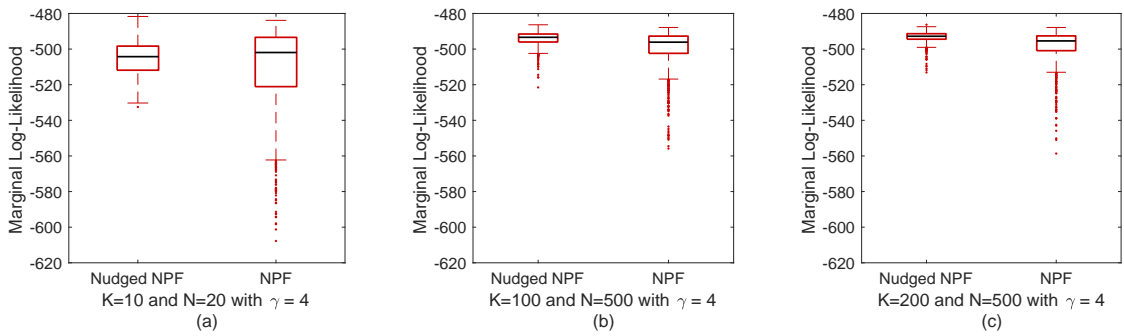}
\end{center}
\caption{Model evidence estimates produced by the nudged NPF and the conventional NPF with varying computational effort. From (a) to (c), it can be seen that, as we increase the number of particles in the parameter space ($K$) and the state space ($N$), the variances of the estimates are smaller. The nudged NPF results in much more stable estimates, with lower variance and fewer extreme values.}
\label{figNestedPF}
\end{figure*}
%

%
In this section, we illustrate the application of the NuPF to estimate the parameters of a financial time-series model. In particular, we adopt a stochastic-volatility SSM and we aim at estimating its unknown parameters (and track its state variables) using the EURUSD log-return data from 2014-12-31 to 2016-12-31 (obtained from \texttt{www.quandl.com}). For this task, we apply two recently proposed Monte Carlo schemes: the nested particle filter (NPF) \citep{Crisan18bernoulli} (a purely recursive, particle-filter style Monte Carlo method) and the particle Metropolis-Hastings (pMH) algorithm \citep{andrieu2010particle} (a batch Markov chain Monte Carlo procedure). In their original forms, both algorithms use the marginal likelihood estimators given by the BPF to construct a Monte Carlo approximation of the posterior distribution of the unknown model parameters. Here, we compare the performance of these algorithms when the marginal likelihoods are computed using either the BPF or the proposed NuPF. 

We assume the stochastic volatility SSM \citep{tsay2005analysis},
\begin{align}
x_0 &\sim \NPDF\left(\mu, \frac{\sigma_v^2}{1-\phi^2}\right), \\
x_t|x_{t-1} &\sim \NPDF(\mu + \phi (x_{t-1} - \mu), \sigma_v^2), \\
y_t|x_t &\sim \NPDF(0,\exp(x_t)),
\end{align}
where $\mu \in \bR$, $\sigma_v \in \bR_+$, and $\phi \in [-1,1]$ are fixed but unknown parameters. The states $\{x_t\}_{1 \leq t \leq T}$ are log-volatilities and the observations $\{y_t\}_{1\leq t \leq T}$ are log-returns. We follow the same procedure as \citet{dahlin2015getting} to pre-process the observations. Given the historical price sequence $s_0, \ldots, s_T$, the log-return at time $t$ is calculated as
\begin{align*}
y_t = 100 \log(s_t/s_{t-1})
\end{align*}
for $1\leq t \leq T$. Then, given $y_{1:T}$, we tackle the joint Bayesian estimation of $x_{1:T}$ and the unknown parameters $\theta = (\mu,\sigma_v,\phi)$. In the next two subsections we compare the conventional BPF and the NuPF as building blocks of the NPF and the pMH algorithms.

\subsection{Nudging the nested particle filter}

The NPF in \citet{Crisan18bernoulli} consists of two layers of particle filters which are used to jointly approximate the posterior distributions of the parameters and the states. The filter in the first layer builds a particle approximation of the marginal posterior distribution of the parameters. Then, for each particle in the parameter space, say $\theta^{(i)}$, there is an \textit{inner} filter that approximates the posterior distribution of the states conditional on the parameter vector $\theta^{(i)}$. The inner filters are classical particle filters, which are essentially used to compute the importance weights (marginal likelihoods) of the particles in the parameter space. In the implementation of \citet{Crisan18bernoulli}, the inner filters are conventional BPFs. We have compared this conventional implementation with an alternative one where the BPFs are replaced by the NuPFs. For a detailed description of the NPF, see \citet{Crisan18bernoulli}.

In order to assess the performances of the nudged and classical versions of the NPF, we compute the model evidence estimate given by the nested filter by integrating out both the parameters and the states. In particular, if the set of particles in the parameter space at time $t$ is $\{ \theta_t^{(i)} \}_{i=1}^K$ and for each particle $\theta_t^{(i)}$ we have a set of particles in the state space $\{x_t^{(i,j)}\}_{j=1}^N$, we compute 
\begin{align*}
\widehat{{\sf p}(y_{1:T})} = \prod_{t=1}^T \left[ \frac{1}{KN} \sum_{i=1}^K \sum_{j=1}^N g_t(x_t^{(i,j)}) \right].
\end{align*}
The model evidence quantifies the fitness of the stochastic volatility model for the given dataset, hence we expect to see a higher value when a method attains a better performance (the intuition is that if we have better estimates of the parameters and the states, then the model will fit better). For this experiment, we compute the model evidence for the nudged NPF {\em before} the nudging step, so as to make the comparison with the conventional algorithm fair.

We have conducted 1,000 independent Monte Carlo runs for each algorithm and computed the model evidence estimates. We have used the same parameters and the same setup for the two versions of the NPF (nudged and conventional). In particular, each unknown parameter is jittered independently. The parameter $\mu$ is jittered with a zero-mean Gaussian kernel variance $\sigma_\mu^2 = 10^{-3}$, the parameter $\sigma_v$ is jittered with a truncated Gaussian kernel on $(0,\infty)$ with variance ${\sigma}_{\sigma_v}^2 = 10^{-4}$, and the parameter $\phi$ is jittered with a zero-mean truncated Gaussian kernel on $[-1,1]$, with variance $\sigma_\phi^2 = 10^{-4}$. We have chosen a large step-size for the nudging step, $\gamma = 4$,  and we have used batch nudging with $M = \lfloor \sqrt{N} \rfloor$.

The results in Fig.~\ref{figNestedPF} demonstrate empirically that the use of the nudging step within the NPF reduces the variance of the model evidence estimators, hence it improves {the numerical stability of the NPF}.

\begin{figure*}[t]
\begin{center}
\includegraphics[scale=0.54]{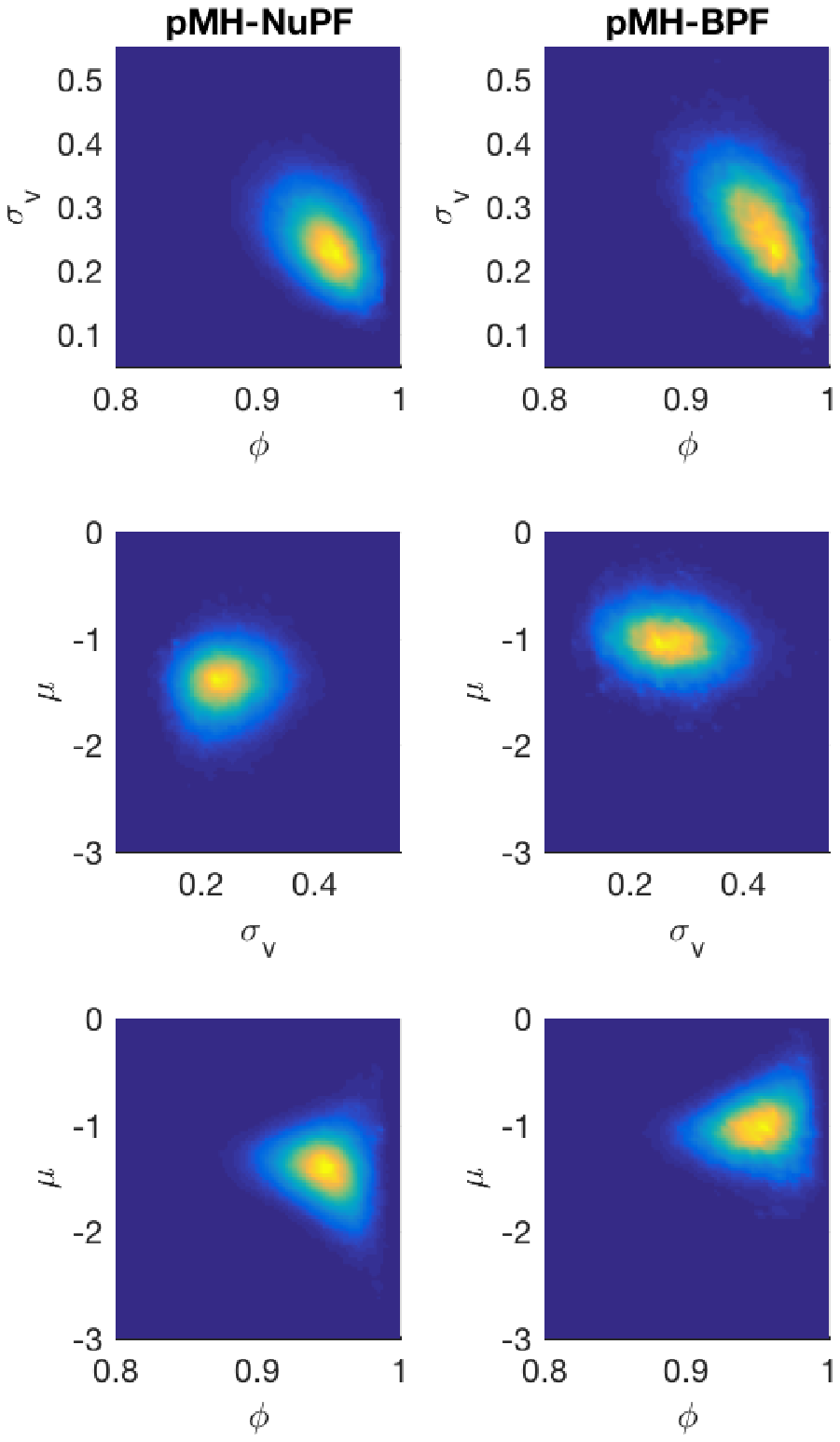}
\includegraphics[scale=0.54]{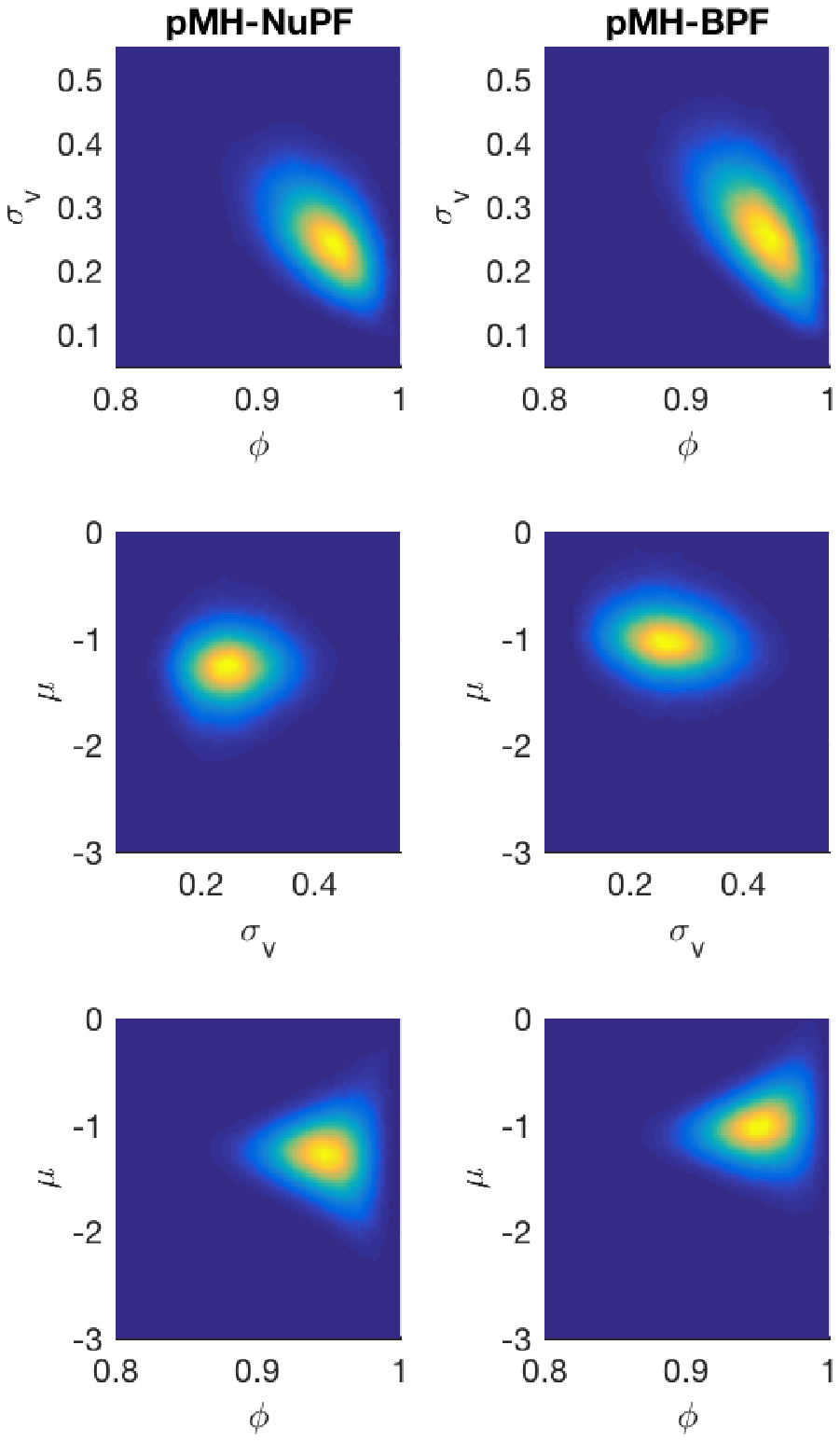}
\includegraphics[scale=0.54]{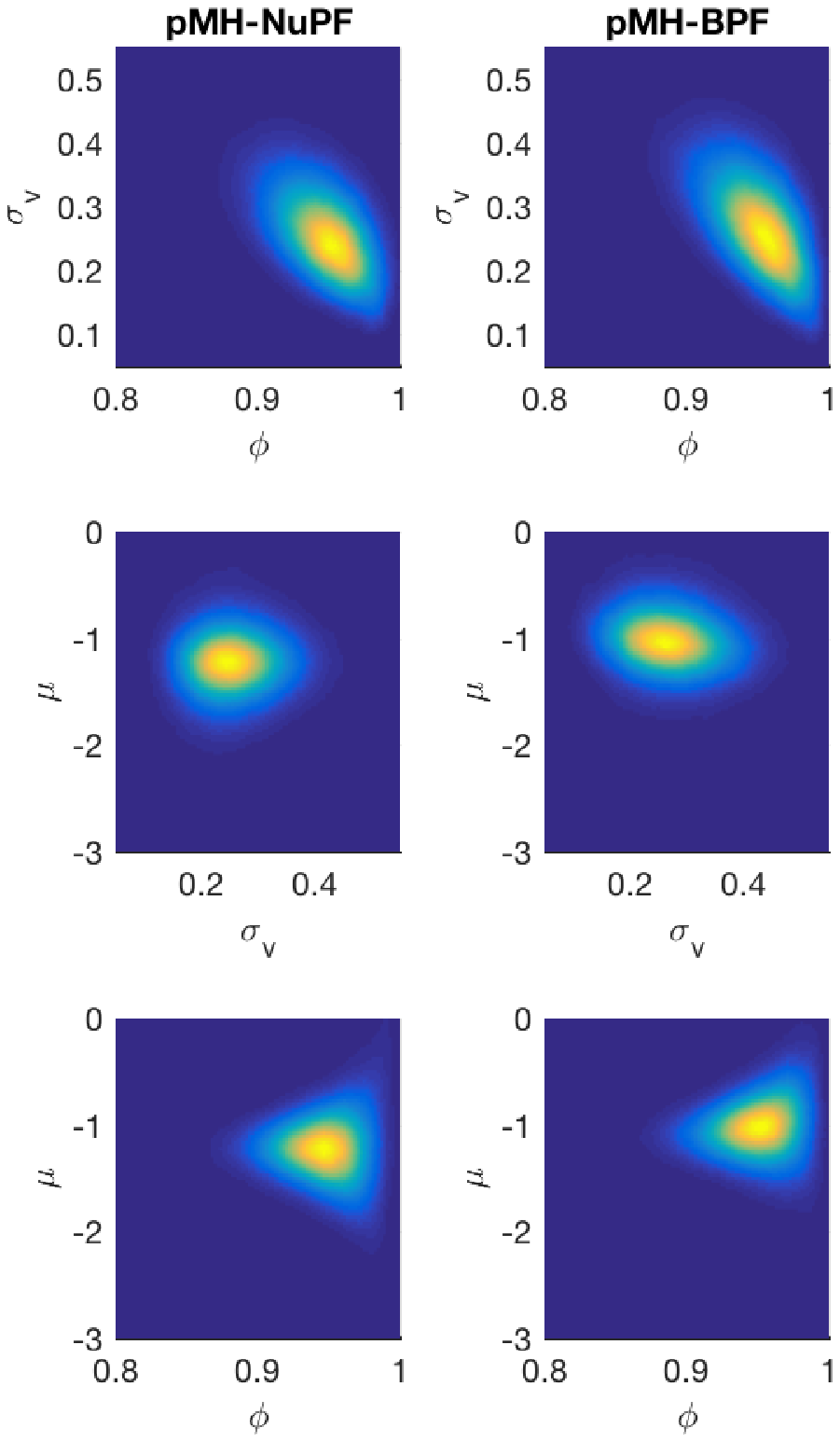}\\
\hspace{0cm}(a) $N = 100$ \hspace{4cm}(b) $N = 500$\hspace{4cm}(c) $N = 1000$
\end{center}
\caption{{The parameter posterior distributions found by the pMH-NuPF and the pMH-BPF for varying $N$. It can be seen that, as $N$ increases, the impact of the nudging-induced bias on the posterior distributions vanishes.}}
\label{figPosteriors}
\end{figure*}

%
\subsection{Nudging the particle Metropolis-Hastings}

The pMH algorithm is a Markov chain Monte Carlo (MCMC) method for inferring parameters of general SSMs \citep{andrieu2010particle}. The pMH uses PFs as auxiliary devices to estimate parameter likelihoods in a similar way as the NPF uses them to compute importance weights. In the case of the pMH, these estimates should be unbiased and they are needed to determine the acceptance probability for each element of the Markov chain. For the details of the algorithm, see \citet{andrieu2010particle} (or \citet{dahlin2015getting} for a tutorial-style introduction). Let us note that the use of NuPF does not lead to an unbiased estimate of the likelihood with respect to the assumed SSM. However, {as discussed in Section \ref{ssModelling}, one can view the use of nudging in this context as an implementation of pMH with an implicit dynamical model $\cM_1$ derived from the original SSM $\cM_0$.}

\begin{figure*}[t]
\begin{center}
\includegraphics[scale=0.43]{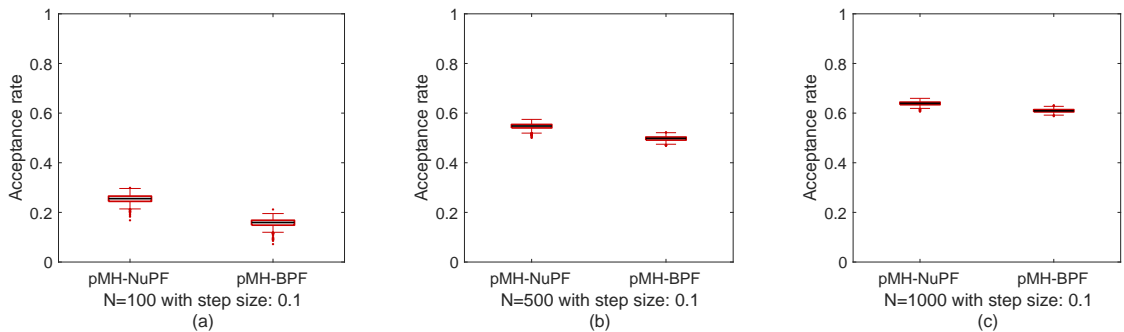}
\end{center}
\caption{Empirical acceptance rates computed for the pMH running BPF and the pMH running NuPF. From (a), it can be seen that there is a significant increase in the acceptance rates when the number of particles are low, e.g., $N = 100$. From (b) and (c), it can be seen that the pMH-NuPF is still better for increasing number of particles but the pMH-BPF is catching up with the performance of the pMH-NuPF.}
\label{figPMH_accRates}
\end{figure*}

We have carried out a computer experiment to compare the performance of the pMH scheme using either BPFs or NuPFs to compute acceptance probabilities. The two algorithms are labeled pMH-BPF and pMH-NuPF, respectively, hereafter. The parameter priors in the experiment are
\begin{align*}
p(\mu) = \NPDF(0,1), \,\,\,\,\,\,\,\, p(\sigma_v) = \mathcal{G}(2,0.1), \,\,\,\,\,\,\,\, p(\phi) = \mathcal{B}(120,2),
\end{align*}
where $\mathcal{G}(a,\theta)$ denotes the Gamma pdf with shape parameter $a$ and scale parameter $\theta$, and $\mathcal{B}(\alpha,\beta)$ denotes the Beta pdf with shape parameters $(\alpha,\beta)$. Unlike \citet{dahlin2015getting}, who use a truncated Gaussian prior centered on $0.95$ with a small variance for $\phi$, we use the Beta pdf, which is defined on $[0,1]$, with mean $\alpha/(\alpha + \beta) = 0.9836$, which puts a significant probability mass on the interval $[0.9,1]$.

We have compared the pMH-BPF algorithm and the pMH-NuPF scheme (using a batch nudging procedure with $\gamma = 0.1$ and $M = \lfloor \sqrt{N}\rfloor$) by running 1,000 independent Monte Carlo trials. We have computed the marginal likelihood estimates in the NuPF {\em after} the nudging step. 

\begin{figure*}[t]
\begin{center}
\includegraphics[scale=0.64]{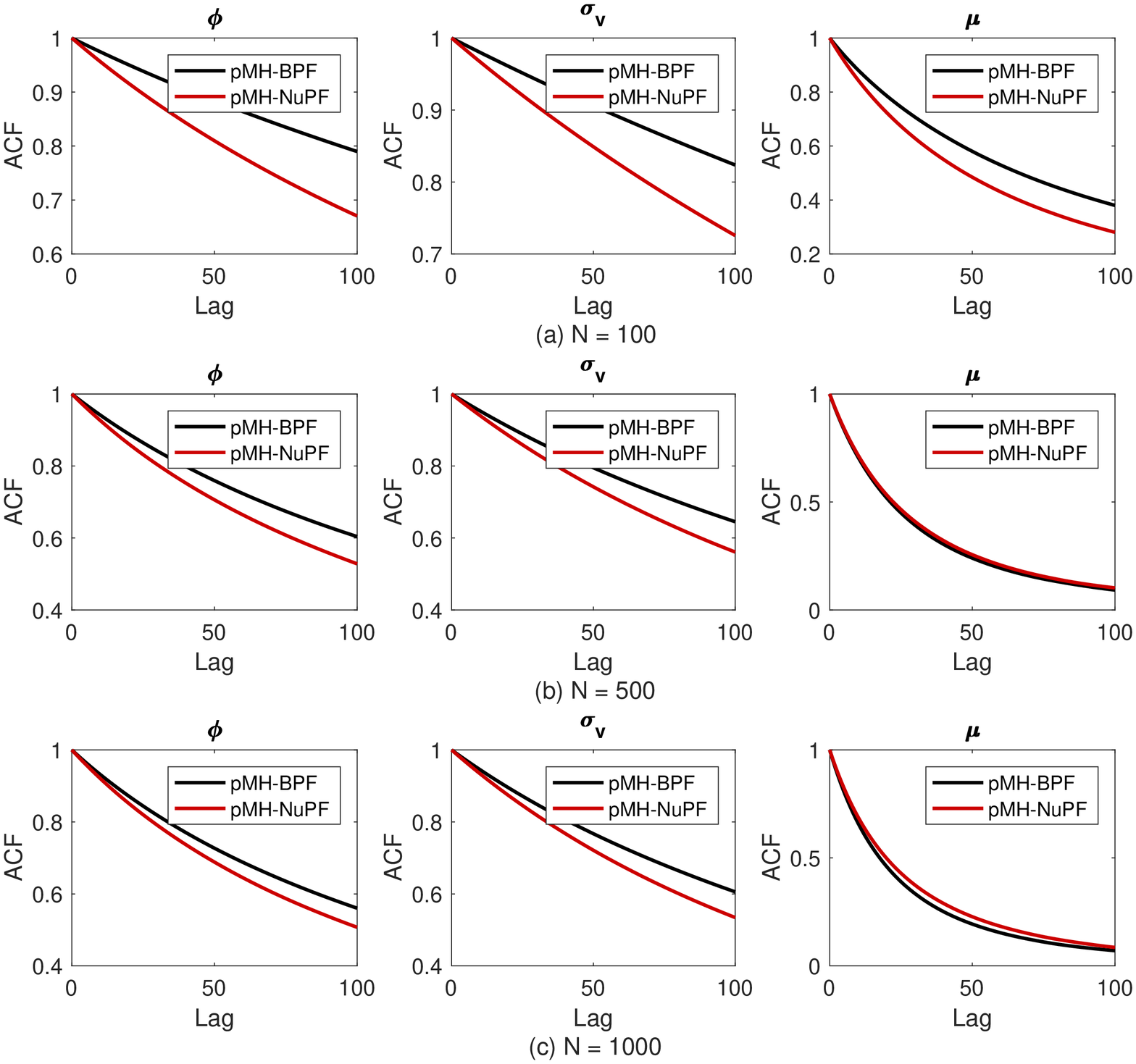}
\end{center}
\caption{Empirical autocorrelation functions (ACFs) computed for the pMH -BPF and the pMH-NuPF. From (a)--(c), it can be seen that using the NuPF instead of BPF within the pMH causes faster autocorrelation decay. These results are obtained by averaging ACFs over $1,000$ Monte Carlo runs.}
\label{figPMH_Autocorr}
\end{figure*}

{First, in order to illustrate the impact of the nudging on the parameter posteriors, we have run the pMH-NuPF and the pMH-BPF and obtained a long Markov chain ($2\times 10^6$ iterations) from both algorithms. Figure \ref{figPosteriors} displays the two-dimensional marginals of the resulting posterior distribution. It can be seen from Fig.~\ref{figPosteriors} that the bias of the NuPF yields a perturbation compared to the posterior distribution approximated with the pMH-BPF. The discrepancy is small but noticeable for small $N$ (see Fig.~\ref{figPosteriors}(a) for $N=100$) and vanishes as we increase $N$ (see Fig.~\ref{figPosteriors}(b) and (c), for $N=500$ and $N=1,000$, respectively). We observe that for a moderate number of particles, such as $N=500$ in Fig.~\ref{figPosteriors}(b), the error in the posterior distribution due to the bias in the NuPF is very slight.}

{Two common figures of merit for MCMC algorithms are the acceptance rate of the Markov kernel (desirably high) and the autocorrelation function of the chain (desirably low). Figure \ref{figPMH_accRates} shows the acceptance rates for the pMH-NuPF and the pMH-BPF algorithms with $N=100$, $N=500$ and $N=1,000$ particles in both PFs. It is observed that the use of nudging leads to noticeably higher acceptance rates, although the difference becomes smaller as $N$ increases.}

{Figure \ref{figPMH_Autocorr} displays the average autocorrelation functions (ACFs) of the chains obtained in the 1,000 independent simulations.We see that the autocorrelation of the chains produced by the pMH-NuPF method decays more quickly than the autocorrelation of the chains output by the conventional pMH-BPF, especially for lower values of $N$. Even for $N=1,000$ (which ensures an almost negligible perturbation of the posterior distribution, as shown in Figure \ref{figPosteriors}(c)) there is an improvement in the ACFs of the parameters $\phi$ and $\sigma_v$ using the NuPF.   Less correlation can be expected to translate into better estimates as well for a fixed length of the chain.}

\section{Conclusions}\label{secConclusions}

We have proposed a simple modification of the particle filter which, according to our computer experiments, can improve the performance of the algorithm (e.g., when tracking high-dimensional systems) or enhance its robustness to model mismatches in the state equation of a SSM. The modification of the standard particle filtering scheme consists of an additional step, which we term nudging, in which a subset of particles are pushed towards regions of the state space with a higher likelihood. In this way, the state space can be explored more efficiently while keeping the computational effort at nearly the same level as in a standard particle filter. We refer to the new algorithm as the ``nudged particle filter'' (NuPF). While, for clarity and simplicity, we have kept the discussion and the numerical comparisons restricted to the modification (nudging) of the conventional BPF, the new step can be naturally incorporated to most known particle filtering methods. 

{We have presented a basic analysis of the NuPF which indicates that the algorithm converges (in $L_p$) with the same error rate as the standard particle filter. In addition, we have also provided a simple reinterpretation of nudging that illustrates why the NuPF tends to outperform the BPF when there is some mismatch in the state equation of the SSM. To be specific, we have shown that, given a fixed sequence of observations, the NuPF amounts to a standard PF for a modified dynamical model which empirically leads to a higher model evidence (i.e., a higher likelihood) compared to the original SSM.}

{The analytical results have been supplemented with a number of computer experiments, both with synthetic and real data.} In the latter case, we have tackled the fitting of a stochastic volatility SSM using Bayesian methods for model inference and a time-series dataset consisting of euro-to-US-dollar exchange rates over a period of two years. We have shown how different figures of merit (model evidence, acceptance probabilities or autocorrelation functions) improve when using the NuPF, instead of a standard BPF, in order to implement a nested particle filter \citep{Crisan18bernoulli} and a particle Metropolis-Hastings \citep{andrieu2010particle} algorithm.
 
Since the nudging step is fairly general, it can be used with a wide range of differentiable or non-differentiable likelihoods. Besides, the new operation does not require any modification of the well-defined steps of the PF so it can be plugged into a variety of common particle filtering methods. Therefore, it can be adopted by a practitioner with hardly any additional effort. In particular, gradient nudging steps (for differentiable log-likelihoods) can be implemented using automatic differentiation tools, currently available in many software packages, hence relieving the user from explicitly calculating the gradient of the likelihood. 

Similar to the resampling step, which is routinely employed for numerical stability, we believe the nudging step can be systematically used for improving the performance and robustness of particle filters.

%

%
\appendix

%

%
\section{Proof of Theorem \ref{ThmNudgingBound}} \label{apNudgingBound}

In order to prove Theorem~\ref{ThmNudgingBound}, we need a preliminary lemma, which can be found, e.g., in \citet{crisan2001particle}.
\begin{lem}\label{lem:Ineq} Let $\alpha,\beta,\bar{\alpha},\bar{\beta}\in\cP(\sX)$ be probability measures and $f,h \in B(\sX)$ be two real bounded functions on $\sX$ such that $(h,\bar{\alpha}) > 0$ and $(h,\bar{\beta})>0$. If the identities,
\begin{align*}
(f,\alpha) = \frac{(fh,\bar{\alpha})}{(h,\bar{\alpha})} \quad \textnormal{and} \quad (f,\beta) = \frac{(fh,\bar{\beta})}{(h,\bar{\beta})}
\end{align*}
hold, then we have,
\begin{align*}
| (f,\alpha)-(f,\beta) | \le \frac{
	1
}{
	(h,\bar \alpha)
} \left|
	(fh,\bar \alpha) - (fh,\bar \beta)
\right| + \frac{
	\| f \|_\infty
}{
	(h,\bar \alpha)
} \left|
	(h,\bar \alpha) - (h,\bar \beta)
\right|. 
\end{align*}
\end{lem}

Now we proceed with the proof of Theorem ~\ref{ThmNudgingBound}. We follow the same kind of induction argument as in, e.g., \citet{DelMoral00} and \citet{Crisan18bernoulli}.

For the base case, i.e. $t = 0$, we draw $x_0^{(i)}$, $i=1, ..., N$, i.i.d. from $\pi_0$ and obtain,
\begin{align*}
\|(\varphi,\pi_0^N) - (\varphi,\pi_0)\|_p = \bE\left[\left|\frac{1}{N}\sum_{i=1}^N \left(\varphi(x_0^{(i)}) - (\varphi,\pi_0)\right)\right|^p\right]^{1/p}.
\end{align*}
We define $S_0^{(i)} = \varphi(x_0^{(i)}) - (\varphi,\pi_0)$ and note that $S_0^{(i)}$, $i = 1,\ldots,N$ are zero-mean and independent random variables. Using the Marcinkiewicz-Zygmund inequality \citep{shiryaev1996}, we arrive at,
\begin{align*}
\bE\left[\left|\frac{1}{N}\sum_{i=1}^N S_0^{(i)} \right|^p\right] &\leq \frac{B_{0,p}}{N^p} \bE\left[\left(\sum_{i=1}^N \left| S_0^{(i)}\right|^2\right)^{\frac{p}{2}}\right] \\
&\leq \frac{B_{0,p}}{N^p} \left(N 4 \|\varphi\|_\infty^2\right)^{\frac{p}{2}},
\end{align*}
where $B_{0,p}$ is a constant independent of $N$ and the last inequality follows from $\left|S_0^{(i)}\right| = \left| \varphi(x_0^{(i)}) - (\varphi,\pi_0)\right| \leq 2 \|\varphi\|_\infty$. Therefore, we have proved that Eq.~\eqref{eqJJ0} holds for the base case,
\begin{align*}
\|(\varphi,\pi_0^N) - (\varphi,\pi_0)\|_p \leq \frac{c_{0,p} \|\varphi\|_\infty}{\sqrt{N}},
\end{align*}
where $c_{0,p} = 2 B_{0,p}^{1/p}$ is a constant independent of $N$.

The induction hypothesis is that, at time $t-1$, 
\begin{equation}
\left\| (\varphi,\pi_{t-1}^N) - (\varphi,\pi_{t-1}) \right\|_p \leq \frac{c_{t-1,p} \|\varphi\|_\infty}{\sqrt{N}}
\nonumber
\end{equation}
for some constant $c_{t-1,p} < \infty$ independent of $N$. 

We start analyzing the predictive measure $\xi_t^N$,
\begin{align*}
\xi_t^N(\mbox{d}x) = \frac{1}{N} \sum_{i = 1}^N \delta_{\bar{x}_t^{(i)}}(\mbox{d}x),
\end{align*}
where $\bar{x}_t^{(i)}$, $i = 1,\ldots,N$ are the particles sampled from the transition kernels $\tau_t^{x_{t-1}^{(i)}}(\mbox{d}x_t) \triangleq \tau_t(\mbox{d} x_t |x_{t-1}^{(i)})$. Since we have $\xi_t = \tau_t \pi_{t-1}$ (see Sec.~1.4), a simple triangle inequality yields,
\begin{align}\label{eq:Proof1:PredIneq}
\left\| (\varphi,\xi_t^N) - (\varphi,\xi_t)\right\|_p &= \left\|(\varphi,\xi_t^N) - (\varphi,\tau_t \pi_{t-1})\right\|_p \nonumber \\
&\leq \left\|(\varphi,\xi_t^N) - (\varphi,\tau_t \pi_{t-1}^N)\right\|_p \\
&+ \left\|(\varphi,\tau_t \pi_{t-1}^N)-(\varphi,\tau_t \pi_{t-1})\right\|_p, \nonumber
\end{align}
where,
\begin{align}
(\varphi,\tau_t \pi_{t-1}^N) = \frac{1}{N} \sum_{i=1}^N (\varphi,\tau_t^{x_{t-1}^{(i)}}).
\end{align}
For the sampling step, we aim at bounding the two terms on the rhs of \eqref{eq:Proof1:PredIneq}.

For the first term, we introduce the $\sigma$-algebra generated by the random variables $x_{0:t}^{(i)}$ and $\bar{x}_{1:t}^{(i)}$, $i = 1,\ldots,N$, denoted $\cF_t = \sigma(x_{0:t}^{(i)},\bar{x}_{1:t}^{(i)},i = 1,\ldots,N)$. Since $\pi_{t-1}^N$ is measurable w.r.t. $\cF_{t-1}$, we can write
\begin{align*}
\bE[(\varphi,\xi_t^N) | \cF_{t-1}] = \frac{1}{N} \sum_{i=1}^N (\varphi,\tau_t^{x_{t-1}^{(i)}}) = (\varphi,\tau_t \pi_{t-1}^N).
\end{align*}
Next, we define the random variables $S_t^{(i)} = \varphi(\bar{x}_t^{(i)}) - (\varphi,\tau_t \pi_{t-1}^N)$ and note that, conditional on $\cF_{t-1}$, $S_t^{(i)}$, $i = 1,\ldots,N$ are zero-mean and independent. Then, the approximation error of $\xi_t^N$ can be written as,
\begin{align*}
\bE[\left| (\varphi,\xi^N_t\right.&\left.) - (\varphi,\tau_t \pi_{t-1}^N)\right|^p | \cF_{t-1}]= \bE\left[ \left| \frac{1}{N} \sum_{i=1}^N S_t^{(i)}  \right|^p \Bigg|	 \cF_{t-1} \right].
\end{align*}
Resorting again to the Marcinkiewicz-Zygmund inequality, we can write,
\begin{align*}
\bE\left[ \left| \frac{1}{N} \sum_{i=1}^N S_t^{(i)}  \right|^p \Bigg|	 \cF_{t-1} \right] \leq \frac{B_{t,p}}{N^p} \bE\left[ \left(\sum_{i=1}^N \left|S_t^{(i)}\right|^2  \right)^{\frac{p}{2}} \Bigg|	 \cF_{t-1} \right],
\end{align*}
where $B_{t,p}< \infty$ is a constant independent of $N$. Moreover, since $\left|S_t^{(i)}\right| = \left|\varphi(\bar{x}_t^{(i)}) - (\varphi,\tau_t \pi_{t-1}^N)\right| \leq 2 \|\varphi\|_\infty$, we have,
\begin{align*}
\bE\left[ \left| \frac{1}{N} \sum_{i=1}^N S_t^{(i)}  \right|^p \Bigg|	 \cF_{t-1} \right] \leq \frac{B_{t,p}}{N^p} \left( N 4 \|\varphi\|_\infty^2\right)^{\frac{p}{2}} = \frac{B_{t,p}}{N^{p/2}} 2^p \|\varphi\|_\infty^p.
\end{align*}
If we take unconditional expectations on both sides of the equation above, then we arrive at
\begin{align}
\|(\varphi,\xi^N_t&) - (\varphi,\tau_t \pi_{t-1}^N)\|_p \leq \frac{c_{1,p} \|\varphi\|_\infty}{\sqrt{N}},
\end{align}
where $c_{1,p} = 2 B_{t,p}^{1/p} < \infty$ is a constant independent of $N$.

To handle the second term in the rhs of \eqref{eq:Proof1:PredIneq}, we define $(\bar{\varphi},\pi_{t-1}) = (\varphi,\tau_t \pi_{t-1})$ where $\bar{\varphi}\in B(\sX)$ and given by,
\begin{align*}
\bar{\varphi}(x) = (\varphi,\tau_t^x).
\end{align*}
We also write $(\bar{\varphi},\pi_{t-1}^N) = (\varphi,\tau_t\pi_{t-1}^N)$. Since $\|\bar{\varphi}\|_\infty \leq \|\varphi\|_\infty$, the induction hypothesis leads,
\begin{align}
\|(\varphi,\tau_t\pi_{t-1}^N) - (\varphi,\tau_t\pi_{t-1})\|_p &= \|(\bar{\varphi},\pi_{t-1}^N) - (\bar{\varphi},\pi_{t-1})\|_p \nonumber \\
&\leq \frac{c_{t-1,p}\|\varphi\|_\infty}{\sqrt{N}} \label{eq:Proof1:PredIneq2},
\end{align}
where $c_{t-1,p}$ is a constant independent of $N$. Combining \eqref{eq:Proof1:PredIneq} and \eqref{eq:Proof1:PredIneq2} yields,
\begin{align}\label{PredBound}
\left\| (\varphi, \xi_t^N) - (\varphi,\xi_t) \right\|_p \leq \frac{c_{1,t,p} \|\varphi\|_\infty}{\sqrt{N}}
\end{align}
where $c_{1,t,p} = c_{t-1,p} + c_{1,p} < \infty$ is a constant independent of $N$.

Next, we have to bound the error between the predictive measure $\xi_t^N$ and the nudged measure $\tilde{\xi}_t^N$. As the sets of samples $\{ \bar x_t^{(i)} \}_{i=1}^N$, used to construct $\xi_t^N$, and $\{ \tilde x_t^{(i)} \}_{i=1}^N$, used to construct $\tilde \xi_t^N$ as shown in \eqref{eqDef00}, differ exactly in $M$ particles, namely $\tilde x_t^{(j_1)}, \ldots, \tilde x_t^{(j_M)}$, where $\{ j_1, \ldots, j_M \} = \mathcal{I}_t$, we readily obtain the relationship
\begin{eqnarray} \label{NudgingBound}
\left\| (\varphi, \xi_t^N) - (\varphi,\tilde{\xi}_t^N) \right\|_p &=& \left\| \frac{1}{N} \sum_{i \in \mathcal{I}_t} \left( \varphi(\bar{x}_t^{(i)}) - \varphi(\tilde{x}_t^{(i)}) \right) \right\|_p \nonumber \\
&\leq& \frac{2\|\varphi\|_\infty M}{N} \nonumber \\
&\leq& \frac{2 \|\varphi\|_\infty}{\sqrt{N}} \label{eqJJ1}
\end{eqnarray}
where the first inequality holds trivially (since $|\varphi(x) - \varphi(x')| \leq 2 \|\varphi\|_\infty$ for every $(x,x') \in \mathsf{X}^2$) and the second inequality follows from the assumption $M \leq \sqrt{N}$. Combining \eqref{PredBound} and \eqref{eqJJ1} we arrive at
\begin{align}\label{FinalNudgingBound}
\left\| (\varphi,\xi_t) - (\varphi,\tilde{\xi}_t^N) \right\|_p \leq \frac{\tilde{c}_{1,t} \|\varphi\|_\infty}{\sqrt{N}},
\end{align}
where the constant $\tilde{c}_{1,t,p} = 2 + c_{1,t,p} < \infty$ is independent of $N$. 

Next, we aim at bounding $\|(\varphi,\pi_t) - (\varphi,\tilde{\pi}_t^N)\|_p$ using \eqref{FinalNudgingBound}. We note that, after the computation of weights, we define the weighted random measure,
\begin{align*}
\tilde{\pi}_t^N = \sum_{i=1}^N w_t^{(i)} \delta_{\tilde{x}_t^{(i)}} \quad \textnormal{where} \quad w_t^{(i)} = \frac{g_t(\tilde{x}_t^{(i)})}{\sum_{i=1}^N g_t(\tilde{x}_t^{(i)})}.
\end{align*}
The integrals computed with respect to the weighted measure $\tilde{\pi}_t^N$ takes the form,
\begin{align}\label{eq:Proof1WeightingRandomMeas}
(\varphi,\tilde{\pi}_t^N) = \frac{(\varphi g_t,\tilde{\xi}^N)}{(g_t,\tilde{\xi}_t^N)}.
\end{align}
On the other hand, using Bayes theorem, integrals with respect to the optimal filter can also be written in a similar form as,
\begin{align}\label{eq:Proof1WeightingOptFilt}
(\varphi,\pi_t) = \frac{(\varphi g_t,{\xi}_t)}{(g_t,{\xi}_t)}.
\end{align}
Using Lemma~\ref{lem:Ineq} together with \eqref{eq:Proof1WeightingRandomMeas} and \eqref{eq:Proof1WeightingOptFilt}, we can readily obtain,
\begin{align}\label{eq:Proof1:weightedBoundPrem}
\left|(\varphi,\tilde{\pi}_t^N) - (\varphi,\pi_t)\right| \leq& \frac{1}{(g_t,\xi_t)} \left(
	\| \varphi \|_\infty \left| (g_t,\xi_t) - (g_t,\xi_t^N)\right| \right. \nonumber \\
	&\left. + \left|(\varphi g_t,\xi_t) - (\varphi g_t,\xi_t^N)\right|\right),
\end{align}
where $(g_t,\xi_t) > 0$ by assumption. Using Minkowski's inequality, we can deduce from \eqref{eq:Proof1:weightedBoundPrem} that
\begin{align}\label{eq:Proof1:weightedBoundPrem2}
\left\|(\varphi,\tilde{\pi}_t^N) - (\varphi,\pi_t)\right\|_p \leq& \frac{1}{(g_t,\xi_t)} \left(
	\| \varphi \|_\infty \left\| (g_t,\xi_t) - (g_t,\xi_t^N)\right\|_p \right. \nonumber \\
	&\left. + \left\|(\varphi g_t,\xi_t) - (\varphi g_t,\xi_t^N)\right\|_p\right).
\end{align}
Noting that we have $\|\varphi g_t\|_\infty \leq \|\varphi\|_\infty \|g_t\|_\infty$, \eqref{FinalNudgingBound} and \eqref{eq:Proof1:weightedBoundPrem2} together yield,
\begin{align}\label{WeightingBound}
\left\| (\varphi,{\pi}_t) - (\varphi, \tilde{\pi}_t^N) \right\|_p \leq \frac{c_{2,t,p} \|\varphi\|_\infty}{\sqrt{N}},
\end{align}
where 
\begin{align*}
c_{2,t,p}= \frac{2 \|g_t\|_\infty \tilde{c}_{1,t,p}}{(g_t, \xi_t)} < \infty
\end{align*}
is a finite constant independent of $N$ (the denominator is positive and the numerator is finite as a consequence of Assumption \ref{BoundedAssumption}).

Finally, the analysis of the multinomial resampling step is also standard. We denote the resampled measure as $\pi_t^N$. Since the random variables which are used to construct $\pi_t^N$ are sampled i.i.d from $\tilde{\pi}_t^N$, the argument for the base case can also be applied here to yield,
\begin{align}\label{ResampleBound}
\left\| (\varphi, \tilde{\pi}_t^N) - (\varphi, \pi_t^N) \right\|_p \leq \frac{c_{3,t,p}\|\varphi\|_\infty}{\sqrt{N}},
\end{align}
where $c_{3,t,p} < \infty$ is a constant independent of $N$. Combining bounds \eqref{WeightingBound} and \eqref{ResampleBound} to obtain the inequality \eqref{eqJJ0}, with $c_{t,p} = c_{2,t,p} + c_{3,t,p} < \infty$, concludes the proof.

%
\section{Proof of Lemma \ref{LemGradientBound}} \label{apLemGradientBound}

Since $\tilde{x}_t^{(i)} = \bar{x}_t^{(i)} + \gamma_t \nabla_{x_t} g_t(\bar{x}^{(i)}_t)$, we readily obtain the relationships
\begin{align}\label{EqGradientBound}
\left| \varphi(\tilde{x}_t^{(i)}) - \varphi(\bar{x}_t^{(i)})\right| &\leq L\left\|\tilde{x}_t^{(i)} - \bar{x}_t^{(i)}\right\|_2 \nonumber\\
&= L \gamma_t \left\Vert\nabla_{x_t} g(\bar x^{(i)}_t)\right\Vert_2 \nonumber\\
&\leq \gamma_t L G_t
\end{align}
where the first inequality follows from the Lipschitz assumption, the identity is due to the implementation of the gradient-nudging step and the second inequality follows from Assumption~\ref{GradientBounded}. Then we bound the error $\| (\varphi, {\xi}_t^N) - (\varphi, \tilde{\xi}_t^N) \|_p$ as 
\begin{eqnarray}
\left\| (\varphi, {\xi}_t^N) - (\varphi,\tilde{\xi}_t^N) \right\|_p &=& \left\| \frac{1}{N} \sum_{i\in \mathcal{I}_t} \left( \varphi(\bar{x}_t^{(i)}) - \varphi(\tilde{x}_t^{(i)}) \right) \right\|_p \nonumber\\
&\leq& \frac{1}{N} \sum_{i\in \mathcal{I}} \left\| \varphi(\bar{x}_t^{(i)}) - \varphi(\tilde{x}_t^{(i)}) \right\|_p \nonumber\\
&\leq& \frac{M}{N} \gamma_t L G_t \label{eqJJ3}
\end{eqnarray}
where the identity is a consequence of the construction of $\mathcal{I}_t$ and we apply Minkowski's inequality, \eqref{EqGradientBound} and the assumption $|\mathcal{I}_t|=M$ to obtain \eqref{eqJJ3}. However, we have assumed that $\sup_{1\le t \le T} \gamma_t M \leq \sqrt{N}$, hence
\begin{align*}
\left\| (\varphi, {\xi}_t^N) - (\varphi, \tilde{\xi}_t^N) \right\|_p \leq \frac{LG_t}{\sqrt{N}}. 
\end{align*}

%
\section{Nudging scheme that increases the model evidence} \label{apEvidence}

Consider the SSM $\cM_0=\{\tau_0,\tau_t,g_t\}$ where the likelihoods $(g_t)_{t\ge 1}$ and the Markov kernels $(\tau_t)_{t\ge 1}$ satisfy the regularity assumptions below.

\begin{assumption} \label{asLipschitzGradient}
The functions $\log g_t(x)$, $t =1, 2, ...$, are differentiable and the gradients $\nabla \log g_t(x)$ are Lipschitz with constant ${\sf L}_t^g < \infty$. To be specific,
\begin{equation}
\| \nabla \log g_t(x) - \nabla \log g_t (x') \|_2 \leq \mathsf{L}_t^g \| x - x'\|_2.
\nonumber
\end{equation}
\end{assumption}

\begin{assumption} \label{asTauLipschitz}
The Markov kernels $\tau_t(\md x|x')$ are absolutely continuous with respect to the Lebesgue measure, hence there are conditional pdf's $m_t(x|x')$ such that $\tau_t(\md x|x') = m_t(x|x') \md x$ for any $x'\in\sX$. Moreover, the log-pdf's $\log m_t(x|x')$ are uniformly Lipschitz in $x'$, i.e., there are non-negative bounded functions $L_t(x)$ such that
\begin{align*}
| \log m_t(x|x') - \log m_t(x|x'') | \leq L_t(x) \| x' - x'' \|_2
\end{align*}
and
$
\mathsf{L}^\tau_t = \sup_{x\in\sX} L_t(x) < \infty.
$
\end{assumption}


For any subset $A \subseteq \sX$, let us introduce the indicator function 
$$
{\bf 1}_A(x) := \left\{
	\begin{array}{ll}
	1, &\mbox{if $x\in A$},\\
	0, &\mbox{otherwise.}\\
	\end{array}
\right.
$$
We construct the nudging operator $\alpha_t^{y_t} : \sX\mapsto\sX$ of the form
\begin{equation}
\alpha_t^{y_t}(x) := \left(
	x + \gamma_t \nabla \log g_t(x)
\right) {\bf 1}_{S_{y_t}}(x) + x {\bf 1}_{\overline{S_{y_t}}}(x),
\label{eqCojoNudge} 
\end{equation}
where $S_{y_t} := \left\{ 
	x \in \sX: \| \nabla \log g_t(x_t) \| \ge 2 \mathsf{L}_t^\tau
\right\}$ (recall that $g_t(x)=g_t(y_t|x)$), $\overline{S_{y_t}} = \sX\backslash S_{y_t}$ is the complement of the set $S_{y_t}$ and $\gamma_t>0$ is small enough to guarantee that $g_t(x+\gamma_t\nabla \log g_t(x)) \ge g_t(x)$. Intuitively, this nudging scheme only takes a gradient step when the slope of the likelihood $g_t$ is sufficient to insure an improvement of the likelihood with a small move of the state $x$.

Assume, for simplicity, that we apply the nudging operator \eqref{eqCojoNudge} to every particle at every time step. The recursive step of the resulting NuPF can be outlined as follows:
\begin{enumerate}
\item For $i=1, ..., N$,
	\begin{enumerate}
	\item draw $\bar x_t^{(i)} \sim \tau_t(\md x_t|x_{t-1}^{(i)})$,
	\item nudge every particle, i.e., $\tilde x_t^{(i)} = \alpha_t^{y_t}(\bar x_t^{(i)})$,
	\item and compute weights $w_t^{(i)} \propto g_t(\tilde x_t^{(i)})$.
	\end{enumerate}
\item Resample to obtain $\{ x_t^{(i)} \}_{i=1, ..., N}$.
\end{enumerate}
The asymptotic convergence of this algorithm can be insured whenever the step sizes $\gamma_t$ are selected small enough to guarantee that $\sup_x \| x - \alpha_t^{y_t}(x)\| \le \frac{1}{\sqrt{N}}$ (this is a consequence of Corollary 1 in \citet{Crisan18bernoulli}). The implicit model for this NuPF is $\cM_1=\{\tau_0,\tilde \tau_t,g_t\}$ where the transition kernel is
\begin{equation}
\tilde \tau_t^{y_t}(\md x_t|x_{t-1}) = \int \delta_{\alpha_t^{y_t}(\bar x)}(\md x_t) \tau_t(\md \bar x|x_{t-1}).
\label{eqDefTauTilde}
\end{equation}
Note that for any integrable function $f:\sX\mapsto\sX$ we have
\begin{eqnarray}
(f,\tilde \tau_t^{y_t})(x_{t-1}) &=& \int \int f(x_t) \delta_{\alpha_t^{y_t}(\bar x)}(\md x_t) \tau_t(\md \bar x | x_{t-1}) \nonumber \\
&=& \int (f \circ \alpha_t^{y_t})(\bar x) \tau_t(\md \bar x | x_{t-1}) \nonumber\\
&=& (f \circ \alpha_t^{y_t}, \tau_t)(x_{t-1}),
\label{eqImportantProperty}
\end{eqnarray}
where $(f \circ \alpha_t^{y_t})(x)=f(\alpha_t^{y_t}(x))$ is the composition of $f$ and $\alpha_t^{y_t}$. In particular, we note that $(g_t,\tilde \tau_t^{y_t})(x)= (g_t\circ\alpha_t^{y_t},\tau_t)(x)$ and $({\bf 1}_\sX,\tilde \tau_t^{y_t})(x) = \int {\bf 1}_{\sX}(\alpha_t^{y_t}(x)) \tau_t(\md \bar x | x) = 1$ because $\alpha_t^{y_t}$ is $\sX\mapsto\sX$ and, therefore, ${\bf 1}_\sX \circ \alpha_t^{g_t} = 1$.

Let $\beta_t(x) := x + \gamma_t \nabla \log g_t(x)$.  Assumptions \ref{asLipschitzGradient} and \ref{asTauLipschitz} entail the following result.


\begin{lem} \label{lmRatios}
If Assumptions \ref{asLipschitzGradient} and \ref{asTauLipschitz} hold and the inequalities
$$
\gamma_t \leq \frac{1}{\mathsf{L}_t^g}
\quad \mbox{and} \quad
\| \nabla \log g_t(x_t) \|_2 \ge 2 \mathsf{L}_t^\tau,
$$ 
are satisfied, then 
\begin{equation}
\frac{
	(g_t \circ \beta_t)(x_t)
}{
	g_t(x_t)
} \geq \sup_{x_{t+1}\in\sX} \frac{
	m_{t+1}( x_{t+1} | x_t)
}{
	m_{t+1}(x_{t+1} | \beta_t(x_t))
}.
\label{eqGoodRatio}
\end{equation}
\end{lem}

\noindent \textbf{Proof.} Assumption \ref{asLipschitzGradient} implies that, for any pair $x,x'\in\sX$,
\begin{align}
\log g_t(x) \geq \log g_t(x') + \left\langle \nabla \log g_t(x'), x - x'\right\rangle - \frac{\mathsf{L}_t^g}{2}\|x - x'\|^2_2
\label{quadLowBnd}
\end{align}
see, e.g., \citet[Lemma~3.4]{bubeck2015convex} for a proof. We can readily use \eqref{quadLowBnd}  to obtain a lower bound for $\log g_t(\beta_t(x_t))$. Indeed,
\begin{eqnarray}
\log g_t(\beta_t(x_t)) &\geq& \log g_t(x_t) + \left\langle \nabla \log g_t(x_t), \gamma_t \nabla \log g_t(x_t)\right\rangle \nonumber \\
&& - \frac{\mathsf{L}_t^g\gamma_t^2}{2}\|\nabla \log g_t(x_t)\|^2_2 \nonumber \\
&=& \log g_t(x_t) + \left(\gamma_t -  \frac{\mathsf{L}_t^g\gamma_t^2}{2}\right) \|\nabla \log g_t(x_t)\|^2_2 \nonumber \\
&\geq& \log g_t(x_t) + \frac{\gamma_t}{2} \|\nabla \log g_t(x_t)\|^2_2 \label{eqJJ00}
\end{eqnarray}
where the last inequality follows from the assumption $\gamma_t \leq \frac{1}{\mathsf{L}_t^g}$. In turn, \eqref{eqJJ00} implies
\begin{equation}
\frac{g_t(\beta_t(x_t))}{g_t(x_t)} \geq \exp\left\{\frac{\gamma_t}{2} \|\nabla\log g_t(x_t)\|^2_2\right\}.
\label{GradRatioLwBnd}
\end{equation}

We now turn to the problem of upper bounding the ratio of transition pdf's. From Assumption \ref{asTauLipschitz} and the definition of $\beta_t(x_t)$ we readily obtain that
\begin{equation}
\log m_{t+1}(x_{t+1} | x_t) - \log m_{t+1}(x_{t+1} | \beta_t(x_t)) \le \mathsf{L}_t^\tau \gamma_t \|\nabla \log g_t(x_t)\|_2
\label{eqJJ11}
\end{equation}
holds for any $x_{t+1}\in\sX$. Taking exponentials on both sides of \eqref{eqJJ11} we arrive at
\begin{align}\label{transitionBound2}
\sup_{x_{t+1}\in\sX} \frac{m_{t+1}(x_{t+1} | x_t)}{m_{t+1}(x_{t+1} | \beta_t(x_t))} \le \exp\left\{ \mathsf{L}_t^\tau \gamma_t \|\nabla \log g_t(x_t)\|_2 \right\}.
\end{align}

If $\| \nabla \log g_t(x_t) \|_2=0$ then expressions \eqref{GradRatioLwBnd} and \eqref{transitionBound2} together readily yield the desired relationship \eqref{eqGoodRatio}.

If $\| \nabla \log g_t(x_t) \|_2 > 0$, then the assumption $\| \nabla \log g_t(x_t) \| \ge 2\mathsf{L}_t^\tau$ implies that
$$
\frac{\gamma_t}{2} \|\nabla\log g_t(x_t)\|_2^2 \ge \gamma_t \mathsf{L}_t^\tau \| \nabla\log g_t(x_t)\|_2
$$
which, together with \eqref{GradRatioLwBnd} and \eqref{transitionBound2}, again yield the desired inequality \eqref{eqGoodRatio}. \qed

Finally, we prove that the evidence in favour of $\cM_1$ is greater than the evidence in favour of  $\cM_0$.

\begin{prop}
Let the nudging scheme be defined as in \eqref{eqCojoNudge}. If Assumptions \ref{asLipschitzGradient} and \ref{asTauLipschitz} hold and the inequality
$$
\gamma_t \leq \frac{1}{\mathsf{L}_t^g}
$$ 
is satisfied for $t=1, \ldots,  T<\infty$, then $\sfp(y_{1:T}|\cM_1) \ge \sfp(y_{1:T}|\cM_0)$.
\end{prop}

\noindent \textbf{Proof.} From the definition of $\tilde \tau_t$ in \eqref{eqDefTauTilde} and the ensuing relationship \eqref{eqImportantProperty}, the evidence of model $\cM_1$ can be readily written down as
\begin{eqnarray}
\sfp(y_{1:T}|\cM_1) &=& \int \cdots \int g_T(\alpha_t^{y_T}(x_T)) \times \nonumber \\
&& \times \prod_{t=1}^{T-1} m_{t+1}(x_{t+1}|\alpha_t^{y_t}(x_t)) g_t(\alpha_t^{y_t}(x_t)) \nonumber \\
&& \times m_1(x_1|x_0)m_0(x_0) \md x_0 \cdots \md x_T.
\end{eqnarray}
It is apparent that $g_T(\alpha_T^{y_T}(x_T)) \ge g_T(x_T)$ for every $x_T$. Moreover, for any $x_t \in \sX$, if $x_t \in S_{y_t}$ then 
$$
\| \nabla \log g_t(x_t) \| \ge 2 \mathsf{L}_t^\tau
\quad \mbox{and} \quad
\alpha_t^{y_t}(x_t) = \beta_t(x_t),
$$ 
hence we can apply Lemma \ref{lmRatios}, which yields 
$$
m_{t+1}(x_{t+1}|\alpha_t^{y_t}(x_t)) g_t(\alpha_t^{y_t}(x_t)) \ge m_{t+1}(x_{t+1}|x_t) g_t(x_t)
$$
for every $x_{t+1}$. Alternatively, if $x_t \notin S_{y_t}$, then $\alpha_t^{y_t}(x_t)=x_t$ and, trivially,
$$
m_{t+1}(x_{t+1}|\alpha_t^{y_t}(x_t)) g_t(\alpha_t^{y_t}(x_t)) = m_{t+1}(x_{t+1}|x_t) g_t(x_t).
$$
Therefore,
\begin{eqnarray}
\sfp(y_{1:T}|\cM_1) &\ge& \int \cdots \int g_T(x_T) \prod_{t=1}^{T-1} m_{t+1}(x_{t+1}|x_t) g_t(x_t) \nonumber \\
&& \times m_1(x_1|x_0) m_0(x_0) \md x_0 \cdots \md x_T \nonumber \\
&=& \sfp(y_{1:T}|\cM_0). \nonumber
\end{eqnarray}
\qed

%

\bibliographystyle{spbasic}
\bibliography{draft}

\begin{thebibliography}{44}
\providecommand{\natexlab}[1]{#1}
\providecommand{\url}[1]{{#1}}
\providecommand{\urlprefix}{URL }
\expandafter\ifx\csname urlstyle\endcsname\relax
  \providecommand{\doi}[1]{DOI~\discretionary{}{}{}#1}\else
  \providecommand{\doi}{DOI~\discretionary{}{}{}\begingroup
  \urlstyle{rm}\Url}\fi
\providecommand{\eprint}[2][]{\url{#2}}

\bibitem[{Ades and van Leeuwen(2013)}]{ades2013exploration}
Ades M, van Leeuwen PJ (2013) An exploration of the equivalent weights particle
  filter. Quarterly Journal of the Royal Meteorological Society
  139(672):820--840

\bibitem[{Ades and van Leeuwen(2015)}]{ades2015equivalent}
Ades M, van Leeuwen PJ (2015) The equivalent-weights particle filter in a
  high-dimensional system. Quarterly Journal of the Royal Meteorological
  Society 141(687):484--503

\bibitem[{Anderson and Moore(1979)}]{Anderson79}
Anderson BDO, Moore JB (1979) Optimal Filtering. Englewood Cliffs

\bibitem[{Andrieu et~al(2010)Andrieu, Doucet, and
  Holenstein}]{andrieu2010particle}
Andrieu C, Doucet A, Holenstein R (2010) Particle {M}arkov chain {M}onte
  {C}arlo methods. Journal of the Royal Statistical Society: Series B
  (Statistical Methodology) 72(3):269--342

\bibitem[{Atkins et~al(2013)Atkins, Morzfeld, and Chorin}]{atkins2013implicit}
Atkins E, Morzfeld M, Chorin AJ (2013) Implicit particle methods and their
  connection with variational data assimilation. Monthly Weather Review
  141(6):1786--1803

\bibitem[{Bain and Crisan(2009)}]{crisan2009fundamentals}
Bain A, Crisan D (2009) Fundamentals of stochastic filtering. Springer

\bibitem[{Bengtsson et~al(2008)Bengtsson, Bickel, and Li}]{bengtsson2008curse}
Bengtsson T, Bickel P, Li B (2008) Curse-of-dimensionality revisited: Collapse
  of the particle filter in very large scale systems. In: Probability and
  statistics: Essays in honor of David A. Freedman, Institute of Mathematical
  Statistics, pp 316--334

\bibitem[{Bernardo and Smith(1994)}]{Bernardo94}
Bernardo JM, Smith AFM (1994) Bayesian Theory. Wiley \& {S}ons

\bibitem[{Bertsekas(2001)}]{BertsekasOptCont1}
Bertsekas DP (2001) Dynamic Programming and Optimal Control, Vol. I. Athena
  Scientific, Belmont, MA

\bibitem[{Bubeck et~al(2015)}]{bubeck2015convex}
Bubeck S, et~al (2015) Convex optimization: Algorithms and complexity.
  Foundations and Trends{\textregistered} in Machine Learning 8(3-4):231--357

\bibitem[{Chopin(2004)}]{chopin2004central}
Chopin N (2004) Central limit theorem for sequential {M}onte {C}arlo methods
  and its application to {B}ayesian inference. The Annals of Statistics
  32(6):2385--2411

\bibitem[{Chorin et~al(2010)Chorin, Morzfeld, and Tu}]{chorin2010implicit}
Chorin A, Morzfeld M, Tu X (2010) Implicit particle filters for data
  assimilation. Communications in Applied Mathematics and Computational Science
  5(2):221--240

\bibitem[{Chorin and Tu(2009)}]{chorin2009implicit}
Chorin AJ, Tu X (2009) Implicit sampling for particle filters. Proceedings of
  the National Academy of Sciences 106(41):17249--17254

\bibitem[{Crisan(2001)}]{crisan2001particle}
Crisan D (2001) Particle filters--a theoretical perspective. In: Doucet A., de
  Freitas N., Gordon N. (eds) Sequential Monte Carlo Methods in Practice.
  Statistics for Engineering and Information Science, Springer, New York, NY,
  pp 17--41

\bibitem[{Crisan and Doucet(2002)}]{Crisan02}
Crisan D, Doucet A (2002) A survey of convergence results on particle
  filtering. IEEE {T}ransactions on {S}ignal {P}rocessing 50(3):736--746

\bibitem[{Crisan and Miguez(2017)}]{Crisan17}
Crisan D, Miguez J (2017) Uniform convergence over time of a nested particle
  filtering scheme for recursive parameter estimation in state--space {M}arkov
  models. Advances in Applied Probability 49(4):1170--1200

\bibitem[{Crisan and Miguez(2018)}]{Crisan18bernoulli}
Crisan D, Miguez J (2018) Nested particle filters for online parameter
  estimation in discrete-time state-space {M}arkov models. Bernoulli
  24(4A):3039--3086

\bibitem[{Dahlin and Sch{\"o}n(2015)}]{dahlin2015getting}
Dahlin J, Sch{\"o}n TB (2015) Getting started with particle
  {M}etropolis-{H}astings for inference in nonlinear dynamical models. arXiv
  1511.01707

\bibitem[{Del~Moral(2004)}]{del2004feynman}
Del~Moral P (2004) {F}eynman-{K}ac Formulae: Genealogical and Interacting
  Particle Systems with Applications. Springer

\bibitem[{Del~Moral and Guionnet(1999)}]{del1999central}
Del~Moral P, Guionnet A (1999) Central limit theorem for nonlinear filtering
  and interacting particle systems. Annals of Applied Probability 9(2):275--297

\bibitem[{Del~Moral and Guionnet(2001)}]{del2001stability}
Del~Moral P, Guionnet A (2001) On the stability of interacting processes with
  applications to filtering and genetic algorithms. Annales de l'Institut Henri
  Poincare (B) Probability and Statistics 37(2):155--194

\bibitem[{{Del Moral} and Miclo(2000)}]{DelMoral00}
{Del Moral} P, Miclo L (2000) Branching and interacting particle systems.
  {A}pproximations of {F}eynman-{K}ac formulae with applications to non-linear
  filtering. In: Az\'{e}ma J., Ledoux M., \'{E}mery M., Yor M. (eds)
  S\'{e}minaire de Probabilit\'{e}s XXXIV. Lecture Notes in Mathematics,
  Springer, Berlin, Heidelberg, vol 1729, pp 1--145

\bibitem[{Douc and Moulines(2008)}]{douc2007limit}
Douc R, Moulines E (2008) Limit theorems for weighted samples with applications
  to sequential {M}onte {C}arlo methods. Annals of Statistics 36(5):2344--2376

\bibitem[{Douc et~al(2009)Douc, Moulines, and Olsson}]{douc2009optimality}
Douc R, Moulines E, Olsson J (2009) Optimality of the auxiliary particle
  filter. Probability and Mathematical Statistics 29(1):1--28

\bibitem[{Doucet et~al(2000)Doucet, Godsill, and
  Andrieu}]{doucet2000sequential}
Doucet A, Godsill S, Andrieu C (2000) On sequential {M}onte {C}arlo sampling
  methods for {B}ayesian filtering. Statistics and {C}omputing 10(3):197--208

\bibitem[{Doucet et~al(2001)Doucet, De~Freitas, and
  Gordon}]{de2001introduction}
Doucet A, De~Freitas N, Gordon N (2001) An introduction to sequential {M}onte
  {C}arlo methods. In: Doucet A., de Freitas N., Gordon N. (eds) Sequential
  Monte Carlo Methods in Practice. Statistics for Engineering and Information
  Science, Springer, New York, NY, pp 3--14

\bibitem[{Gordon et~al(1993)Gordon, Salmond, and Smith}]{gordon1993novel}
Gordon NJ, Salmond DJ, Smith AF (1993) Novel approach to
  nonlinear/non-{G}aussian {B}ayesian state estimation. In: IEE Proceedings F
  (Radar and Signal Processing), IET, vol 140, pp 107--113

\bibitem[{Hoke and Anthes(1976)}]{hoke1976initialization}
Hoke JE, Anthes RA (1976) The initialization of numerical models by a
  dynamic-initialization technique. Monthly Weather Review 104(12):1551--1556

\bibitem[{Johansen and Doucet(2008)}]{johansen2008note}
Johansen AM, Doucet A (2008) A note on auxiliary particle filters. Statistics
  \& {P}robability {L}etters 78(12):1498--1504

\bibitem[{Kitagawa(1996)}]{kitagawa1996monte}
Kitagawa G (1996) {M}onte {C}arlo filter and smoother for non-{G}aussian
  nonlinear state space models. Journal of {C}omputational and {G}raphical
  {S}tatistics 5(1):1--25

\bibitem[{K{\"u}nsch(2005)}]{kunsch2005recursive}
K{\"u}nsch HR (2005) Recursive {M}onte {C}arlo filters: algorithms and
  theoretical analysis. Annals of Statistics 33(5):1983--2021

\bibitem[{van Leeuwen(2009)}]{van2009particle}
van Leeuwen PJ (2009) Particle filtering in geophysical systems. Monthly
  Weather Review 137(12):4089--4114

\bibitem[{van Leeuwen(2010)}]{van2010nonlinear}
van Leeuwen PJ (2010) Nonlinear data assimilation in geosciences: an extremely
  efficient particle filter. Quarterly Journal of the Royal Meteorological
  Society 136(653):1991--1999

\bibitem[{Liu and Chen(1998)}]{liu1998sequential}
Liu JS, Chen R (1998) Sequential {M}onte {C}arlo methods for dynamic systems.
  Journal of the {A}merican {S}tatistical {A}ssociation 93(443):1032--1044

\bibitem[{Malanotte-Rizzoli and Holland(1986)}]{malanotte1986data}
Malanotte-Rizzoli P, Holland WR (1986) Data constraints applied to models of
  the ocean general circulation. {P}art {I}: {T}he steady case. Journal of
  {P}hysical {O}ceanography 16(10):1665--1682

\bibitem[{Malanotte-Rizzoli and Holland(1988)}]{malanotte1988data}
Malanotte-Rizzoli P, Holland WR (1988) Data constraints applied to models of
  the ocean general circulation. {P}art {II}: the transient, eddy-resolving
  case. Journal of {P}hysical {O}ceanography 18(8):1093--1107

\bibitem[{M{\'\i}guez et~al(2013)M{\'\i}guez, Crisan, and
  Djuri{\'c}}]{miguez2013convergence}
M{\'\i}guez J, Crisan D, Djuri{\'c} PM (2013) On the convergence of two
  sequential {M}onte {C}arlo methods for maximum a posteriori sequence
  estimation and stochastic global optimization. Statistics and Computing
  23(1):91--107

\bibitem[{Oreshkin and Coates(2011)}]{Oreshkin11}
Oreshkin BN, Coates MJ (2011) Analysis of error propagation in particle filters
  with approximation. The Annals of Applied Probability 21(6):2343--2378

\bibitem[{Pitt and Shephard(1999)}]{pitt1999filtering}
Pitt MK, Shephard N (1999) Filtering via simulation: Auxiliary particle
  filters. Journal of the {A}merican {S}tatistical {A}ssociation
  94(446):590--599

\bibitem[{Robert(2007)}]{Robert07}
Robert CP (2007) The {B}ayesian Choice. Springer

\bibitem[{Shiryaev(1996)}]{shiryaev1996}
Shiryaev AN (1996) Probability. Springer

\bibitem[{Snyder et~al(2008)Snyder, Bengtsson, Bickel, and
  Anderson}]{snyder2008obstacles}
Snyder C, Bengtsson T, Bickel P, Anderson J (2008) Obstacles to
  high-dimensional particle filtering. Monthly Weather Review
  136(12):4629--4640

\bibitem[{Tsay(2005)}]{tsay2005analysis}
Tsay RS (2005) Analysis of {F}inancial {T}ime {S}eries. John Wiley \& Sons

\bibitem[{Zou et~al(1992)Zou, Navon, and LeDimet}]{zou1992optimal}
Zou X, Navon I, LeDimet F (1992) An optimal nudging data assimilation scheme
  using parameter estimation. Quarterly Journal of the Royal Meteorological
  Society 118(508):1163--1186

\end{thebibliography}



\end{document}